# Extended validations on photon number resolving detector based Gaussian boson sampling with low noises

Yang Ji[1,2], Yongzheng Wu[1,2], Shi Wang[1,2], Jie Hou[1,2], Zijian Wang[1], Bo Jiang[*1]
[1]*The 32nd Research Institute of China Electronics Technology Group Corporation, Shanghai 201808, China*
[2]*Shanghai Research Center for Quantum Sciences, Shanghai 201315, China*
*Corresponding author. E-mail: b26jiang@126.com

## Abstract

Gaussian boson sampling (GBS) is a variety of boson sampling overcoming the stable single-photon preparation difficulty of the later. However, like those in the original version, noises in GBS will also result in the deviation of output patterns and the reduction of classical simulation complexity. We extend the pattern recognition validation, together with the correlation approach as a comparison, on GBS using photon number resolving detectors, with noises of both photon loss and distinguishability, to quantificationally evaluate noise levels. As for the classical simulation with noises to be used during validations, it is actually a simulation of mixed states where we employ an existing photon-pair strategy to realize polynomial speedup locally. Furthermore, we use an output-binning strategy to realize validation speedup. Our simulation indicates that the pattern recognition protocol is useful for noise evaluations of GBS even when noises are sufficiently low.

## 1 Introduction

Boson sampling proposed by Aaronson and Arkhipov (AABS) is a quantum computing candidate to demonstrate the so-called quantum computational advantage [1]. This proposal is found to be hard to realize in a large scale due to the scarce of indistinguishable single-photon sources [2-4]. Gaussian boson sampling (GBS) is a substituted scheme where squeezed vacuum states are used as input sources, which are believed to be easier to prepare [5, 6]. These states enter into a passive linear interferometer and output photons are collected by photon number resolving detectors (PNRDs) [7] or threshold detectors [8]. In the noiseless assumption, the output states would be Gaussian because the input states are naturally Gaussian and the transformation between them are unitary. The output probability can be obtained through computing the hafnian of a symmetric matrix, if the displacement of input states is zero. This is thought to be classically #-P hard as that of the permanent corresponding to AABS. GBS is developed physically, reaching a stage where the plausible quantum computational advantage has been achieved [9-12] and some implements applying GBS are available meanwhile [13-19].

However, noises exist physically, such as photon distinguishability resulted from intrinsic differences of photons with imperfect light sources [20-23], photon loss during propagations [24-27], and dark counts occurring in detectors [28]. These noises will not only cause output deviations but also make practical GBS easier to be classically simulated [23-26]. Therefore, with noises, claims of the quantum computational advantage may be unreliable, even though the output photon number may be large. Indeed, output photons tend to come from classical processes instead of multiphoton interferences, when the system is noisy enough. On this condition, it is up to those validation approaches [29-37] to quantificationally evaluate whether the actual noise level is too high to meet the advantage requirement. This is stricter compared to those tasks of just discriminating GBS (noisy or not) from classically easier mockups in a binary way.

There are several validation approaches, some designed for GBS typically. The Bayesian approach [29, 30] is universal but needs large amounts of hafnian (or loop-hafnian) calculations, only suitable to small-scale experiments. The low-order correlation approach [31, 32] focusing on output photon number correlations in modes, is computationally efficient but insensitive to high order multiphoton interferences, and thus, may be not accurate to deal with the noise of photon distinguishability. Nevertheless, with the 2nd correlators, photon loss and the convoluted noise contribution can be recognized [33]. The approach based on the binned output photon number distribution differences between experiments and some other mockups to exclude, is also efficient because the binned photon number distribution can be approximated in polynomial time, providing a semiquantitative evaluation [34, 35]. Despite in AABS, the output-binning approach shows a response to photon distinguishability, which is coarsely revealed by the width of a Gaussian curve [34], it is still not clear how accurate will this approach be, if it is used when the noises are sufficiently low but are still enough to be employed in efficient classical simulations, especially in GBS. Another efficient approach based on the graph theory, depicts an orbit for each kind of samples in a three-dimensional coordinate space of output patterns [36]. Recently, this approach is found to be effective in time-bin GBS to qualitatively evaluate the level of photon loss [37], whereas it may only work in the regime where the output photon number is much smaller than the mode number.

Here, we extend the validations on noisy GBS based on pattern recognition techniques [38]. More precisely, we compare pattern differences correlated to both photon loss and photon distinguishability, at noise levels quite low that the accuracy of validation approaches is challenged. It has been shown to be feasible with an extension on noisy AABS in the collision-free regime, where the pattern recognition approach is found to be sensitive to photon distinguishability, especially when photons are close to be indistinguishable and thus high order multiphoton interferences become significant [39]. This is guaranteed by the fact that the intrinsic data structure of AABS is quite unbalanced and the unbalance will vanish quickly when photon distinguishability is further introduced. While the main deference between AABS and GBS lies in light sources, it is notable that correspondingly the permanent (of a matrix $\boldsymbol{G}$) and the hafnian can be connected by [5]

$$\mathrm{perm}(\boldsymbol{G}) = \mathrm{haf}\left(\begin{bmatrix} \boldsymbol{0} & \boldsymbol{G} \\ \boldsymbol{G}^{\mathrm{t}} & \boldsymbol{0} \end{bmatrix}\right). \tag{1}$$

This indicates that the permanent is a particular case of the hafnian. Indeed, the permanent counts perfect matches in a bipartite graph [40], which is a particular weighted graph where perfect matches are counted by the hafnian. Therefore, since the unbalance in AABS seems to be compatible to Haar random at least, it may reserve in GBS more generally. Let us look at the data structure of noiseless GBS briefly. The output probability tends to be exponentially low when the photon number in each mode increases heavily, indicating that the unbalance remains locally in terms of both probability distributions and norm distances of outputs. Hence, the unbalance looks like to be more natural in GBS.

This paper is constructed as follows. In Sec. 2 we briefly review some basic concepts of GBS. In Sec. 3 we simulate the output patterns based on chain rules, employing an existing photon-pair strategy to locally realize polynomial speedup. In Sec. 4 we extend the pattern recognition validation approach on lossy GBS and give numerical analysis of the data structures. In Sec. 5 we discuss the case with photon distinguishability. In Sec. 6 we use the output-binning strategy to realize validation speedup. In Sec. 7 we perform an extension of the correlation validation approach as a comparison. Lastly, we give a conclusion.

# 2 Gaussian boson sampling

GBS is a specialized quantum computing model, where generally $K$ single-mode squeezed states (SMSSs) enter into a passive linear interferometer characterized by a $m$-dimensional unitary matrix $\boldsymbol{T}$ $(m \geq K)$. $\boldsymbol{T}$ should be random in the Haar measure to meet the complexity requirement of GBS, where optical units like beam splitters and phase shifters can be arrayed within rules [41, 42]. Output photons are collected by PNRDs where several photons are allowed to be measured in each mode, or by threshold detectors where correspondingly at most one click is allowed. No matter what kind of detectors is used, the output patterns represented by Fock states are proved to be hard to simulate with a classical computer, exactly or even approximately [5, 8]. We focus on the PNRD case because it is implemental and also more suitable to utilize pattern recognition validations with a much larger Hilbert space.

In this paper we use the convention that $\hbar = 2$. In the ordering of $(\hat{\boldsymbol{x}}, \hat{\boldsymbol{p}}) = (\hat{x}_1, \dots, \hat{x}_m, \hat{p}_1, \dots, \hat{p}_m)$ where $\hat{x}_i$ and $\hat{p}_i$ are canonical position and momentum operators in the $i$th mode, the ideal covariance matrix of the input state can be expressed as [5]

$$\boldsymbol{V}_{\text{in}} = [(\oplus_{i=1}^{K} e^{2r_i}) \oplus \boldsymbol{I}_{m-K}] \oplus [(\oplus_{i=1}^{K} e^{-2r_i}) \oplus \boldsymbol{I}_{m-K}], \tag{2}$$

where $\boldsymbol{I}_{m-K}$ is the ($m$-$K$)-dimensional identity matrix. The squeezing parameter $r_i$ is set as a constant $r$ in this paper. In the ordering of $(\hat{\boldsymbol{x}} + i\hat{\boldsymbol{p}}, \hat{\boldsymbol{x}} - i\hat{\boldsymbol{p}})$, the covariance matrix of the Q-function is [8]

$$\boldsymbol{Q}_{\text{in}} = \frac{1}{4}\begin{bmatrix} \boldsymbol{I}_m & i\boldsymbol{I}_m \\ \boldsymbol{I}_m & -i\boldsymbol{I}_m \end{bmatrix} \boldsymbol{V}_{\text{in}} \begin{bmatrix} \boldsymbol{I}_m & i\boldsymbol{I}_m \\ \boldsymbol{I}_m & -i\boldsymbol{I}_m \end{bmatrix}^{\dagger} + \frac{1}{2}\boldsymbol{I}_{2m}. \tag{3}$$

The corresponding matrix of the output state is [22]

$$\boldsymbol{Q}_{\text{out}} = \begin{bmatrix} \boldsymbol{T} & \boldsymbol{0} \\ \boldsymbol{0} & \boldsymbol{T}^{*} \end{bmatrix} \boldsymbol{Q}_{\text{in}} \begin{bmatrix} \boldsymbol{T}^{\dagger} & \boldsymbol{0} \\ \boldsymbol{0} & \boldsymbol{T}^{\text{t}} \end{bmatrix}, \tag{4}$$

where a unitary transformation $\hat{\boldsymbol{T}}$ is performed on the input Gaussian state and the ideal output state is still Gaussian. The kernel matrix of the output state is [5]

$$\boldsymbol{A} = \begin{bmatrix} \boldsymbol{0} & \boldsymbol{I}_m \\ \boldsymbol{I}_m & \boldsymbol{0} \end{bmatrix} (\boldsymbol{I}_{2m} - \boldsymbol{Q}_{\text{out}}^{-1}). \tag{5}$$

Finally, if the displacement of the Gaussian state is zero, the probability of the output pattern $\boldsymbol{s} = (s_1, \dots, s_m)$, represented by the Fock state, can be obtained via [5]

$$\text{pr}(\boldsymbol{s}) = \frac{\text{haf}(\boldsymbol{A}_{\boldsymbol{s}})}{s_1! \dots s_m! \sqrt{\det(\boldsymbol{Q}_{\text{out}})}}. \tag{6}$$

$\boldsymbol{A}_{\boldsymbol{s}}$ is a submatrix of $\boldsymbol{A}$ where the $i$th and the $i + m$-th rows and columns are repeated for $s_i$ times and $\det(\cdot)$ denotes the matrix determinant which can be classically computed in polynomial time. The complexity of GBS relies on the calculation of the hafnian, whereas it can be reduced obviously if the output state is pure. On this condition, $\boldsymbol{A}_{\boldsymbol{s}}$ can be expressed as $\boldsymbol{A}_{\boldsymbol{s}} = \boldsymbol{B}_{\boldsymbol{s}} \oplus \boldsymbol{B}_{\boldsymbol{s}}^{*}$ where generally $\boldsymbol{B}_{\boldsymbol{s}}$ is a matrix with the dimension only half of that of $\boldsymbol{A}_{\boldsymbol{s}}$. As a result, the probability computation can be greatly simplified considering the fact that [5]

$$\text{haf}(\boldsymbol{A}_{\boldsymbol{s}}) = |\text{haf}(\boldsymbol{B}_{\boldsymbol{s}})|^2. \tag{7}$$

In addition, if the displacement $\boldsymbol{R}_{\text{out}} \neq \boldsymbol{0}$, then the complexity relies on the calculation of the loop-hafnian. In this case, the output probability is [43, 44]

$$\text{pr}(\boldsymbol{s}) = \frac{\exp\left(-(1/2)\overline{\boldsymbol{\alpha}}^{\dagger}\boldsymbol{Q}_{\text{out}}^{-1}\overline{\boldsymbol{\alpha}}\right)\text{lhaf}(\text{filldiag}(\boldsymbol{A}_{\boldsymbol{s}},\overline{\boldsymbol{\gamma}}_{\boldsymbol{s}}))}{s_1!\ldots s_m!\sqrt{\det\left(\boldsymbol{Q}_{\text{out}}\right)}}, \tag{8}$$

where $\overline{\boldsymbol{\alpha}} = ((\overline{\boldsymbol{x}} + i\overline{\boldsymbol{p}})/2, (\overline{\boldsymbol{x}} - i\overline{\boldsymbol{p}})/2)$ represents the displacement, $\text{lhaf}(\cdot)$ denotes the loop-hafnian of a matrix, $\overline{\boldsymbol{\gamma}}_{\boldsymbol{s}} = (\overline{\boldsymbol{\beta}}_{\boldsymbol{s}}, \overline{\boldsymbol{\beta}}_{\boldsymbol{s}}^{*})$ is a vector where the $i$th and $i + m$-th elements of $\overline{\boldsymbol{\gamma}} = (\boldsymbol{Q}_{\text{out}}^{-1}\overline{\boldsymbol{\alpha}})^{*} = (\overline{\boldsymbol{\beta}}, \overline{\boldsymbol{\beta}}^{*})$ are repeated for $s_i$ times, and $\text{filldiag}(\boldsymbol{A}_{\boldsymbol{s}}, \overline{\boldsymbol{\gamma}}_{\boldsymbol{s}})$ denotes a variation of $\boldsymbol{A}_{\boldsymbol{s}}$ where the diagonal elements are replaced by those of $\overline{\boldsymbol{\gamma}}_{\boldsymbol{s}}$. If the output state is still pure, the computation can be simplified by [43]

$$\text{lhaf}(\text{filldiag}(\boldsymbol{A}_{\boldsymbol{s}}, \overline{\boldsymbol{\gamma}}_{\boldsymbol{s}})) = |\text{lhaf}(\text{filldiag}(\boldsymbol{B}_{\boldsymbol{s}}, \overline{\boldsymbol{\beta}}_{\boldsymbol{s}}))|^2. \tag{9}$$

The non-zero displacement case is essential, in the simulation of noisy GBS using chain rules [43, 45], where generally conditional states are generated with unfixed displacements.

# 3 Exact simulations of noisy GBS

We exactly simulate noisy GBS based on the chain rule method, employing a photon-pair strategy proposed by Jacob F. F. Bulmer et al. [44] to realize polynomial speedup for output patterns with collision photons.

As for an output state considering noises such as photon loss and photon distinguishability, or a mixed state in other words, it is crucial to use a series of auxiliary variables $\boldsymbol{\alpha} = (\alpha_2, \ldots, \alpha_m)$ from heterodyne measurements [43] to realize the simplification by Eq. (9). In each step of the chain rule method, we face up with a conditional pure state easier to deal with, where the displacement is not fixed. This is feasible entirely considering the fact that once $\boldsymbol{\alpha}$ is obtained from a multivariate Gaussian distribution, a sample could then be generated via

$$\text{pr}(\boldsymbol{s}) \propto \prod_{i=1}^{m} \text{pr}(s_i | s_1, \ldots, s_{i-1}, \alpha_{i+1}, \ldots, \alpha_m). \tag{10}$$

During the simulation, we use the photon-pair strategy, where photons in PNRDs are rearranged with a greedy algorithm [44] to form pairs. Correspondingly, $\text{lhaf}(\text{filldiag}(\boldsymbol{B}_{\boldsymbol{s}}, \overline{\boldsymbol{\beta}}_{\boldsymbol{s}}))$ in Eq. (9) should also be replaced by the function $\text{lhafmix}(\boldsymbol{C}, \overline{\boldsymbol{\mu}}, \boldsymbol{n})$, where the speedup is realized with [44, 46, 47]

$$\text{lhafmix}(\boldsymbol{C}, \overline{\boldsymbol{\mu}}, \boldsymbol{n}) = \frac{1}{2^{N/2}(N/2)!}\sum_{\boldsymbol{z}}\left\{(-1)^{\sum_i (n_i - z_i)/2}\left[\prod_i \binom{n_i}{(n_i+z_i)/2}\right]\frac{d^{N/2}}{d\lambda^{N/2}}\text{f}(\lambda, \boldsymbol{C}_{\boldsymbol{n}}, \overline{\boldsymbol{\mu}}_{\boldsymbol{n}}, \boldsymbol{z})|_{\lambda=0}\right\}, \tag{11}$$

$$\text{f}(\lambda, \boldsymbol{C}_{\boldsymbol{n}}, \overline{\boldsymbol{\mu}}_{\boldsymbol{n}}, \boldsymbol{z}) = \sum_{j=1}^{N/2}\left\{\frac{1}{j!}\left[\sum_{k=1}^{N/2}\left(\frac{\text{tr}((\boldsymbol{C}_{\boldsymbol{n}}\boldsymbol{X}_{\boldsymbol{z}})^k)}{2k} + \frac{\overline{\boldsymbol{\mu}}_{\boldsymbol{n}}\boldsymbol{X}_{\boldsymbol{z}}(\boldsymbol{C}_{\boldsymbol{n}}\boldsymbol{X}_{\boldsymbol{z}})^{k-1}\overline{\boldsymbol{\mu}}_{\boldsymbol{n}}^{\text{t}}}{2}\right)\lambda^k\right]^j\right\}. \tag{12}$$

In Eq. (11) the finite difference sieve method is used to improve accuracies. $\boldsymbol{n}$ is a vector listing the photon pair numbers according to the greedy algorithm, with the total photon number $N = 2\sum_i n_i$. $\boldsymbol{z}$ runs through all the vectors with $z_i \in \{-n_i, -n_i + 2, \ldots, n_i - 2, n_i\}$. Hence the total number of $\boldsymbol{z}$ is $\prod_i (n_i + 1)$. Here the double computation speed can be achieved with noticing that those vectors and corresponding opposite vectors contribute the same in Eq. (11). $\boldsymbol{C}$ and $\overline{\boldsymbol{\mu}}$ contain the information of photon pairs. As for the $k$th pair $(i, j)$, $\boldsymbol{C}$ keeps the $i$th and $j$th rows and columns of $\boldsymbol{B}$ as its $k$th and $k + |\boldsymbol{n}|$-th rows and columns, and $\overline{\boldsymbol{\mu}}$ keeps the $i$th and $j$th elements of $\overline{\boldsymbol{\beta}}$ as its $k$th and $k + |\boldsymbol{n}|$-th elements, where $|\boldsymbol{n}|$ represents the length of $\boldsymbol{n}$. With these operations, both the dimension of $\boldsymbol{C}$ and the length of $\overline{\boldsymbol{\mu}}$ are $2|\boldsymbol{n}|$. $\boldsymbol{C}_{\boldsymbol{n}}$ is a submatrix of $\boldsymbol{C}$, which is constructed like $\boldsymbol{A}_{\boldsymbol{s}}$. $\overline{\boldsymbol{\mu}}_{\boldsymbol{n}}$ is a variation of $\overline{\boldsymbol{\mu}}$ where the $i$th and $i + |\boldsymbol{n}|$-th elements are repeated for $n_i$ times respectively. $\text{tr}(\cdot)$ denotes the matrix trace. Lastly, $\boldsymbol{X}_{\boldsymbol{z}}$ is expressed as

$$\boldsymbol{X_z} = \begin{bmatrix} \boldsymbol{0} & \mathrm{diag}(\boldsymbol{\Delta_z}) \\ \mathrm{diag}(\boldsymbol{\Delta_z}) & \boldsymbol{0} \end{bmatrix}, \tag{13}$$

where $\mathrm{diag}(\cdot)$ denotes the diagonal matrix. $\boldsymbol{\Delta_z}$ is a vector which can be written as $(\boldsymbol{\delta}_1, \ldots, \boldsymbol{\delta}_{|\boldsymbol{n}|})$. In the array $\boldsymbol{\delta}_i$ of which the length is $n_i$, the first $(n_i + z_i)/2$ elements are 1 while the rest $(n_i - z_i)/2$ elements are $-1$.

Some mathematical tools can be used, such as the Schur decomposition when calculating the trace of a multiple matrix product in Eq. (12), to realize further speedup [46]. Some other notices include the unpaired photon case, where a single photon is left after the execution of the greedy algorithm. The photon-pair strategy is still feasible. We intentionally pair the single photon to a virtual one in an additional $m + 1$-th mode [44], considering the fact that for any symmetric matrix $\boldsymbol{G}$, it has $\mathrm{lhaf}(\boldsymbol{G}) \equiv \mathrm{lhaf}(\boldsymbol{G} \oplus 1)$.

# 4 Extended validations on lossy GBS

With methods mentioned above, we can now simulate noisy GBS exactly on a computer, where the zero states have been excluded. We first focus on photon loss. Suppose the transmission rate in each mode is equally $\eta_{\mathrm{t}}$ with balanced loss, the covariance matrix of the output state is [25]

$$\boldsymbol{V}_{\mathrm{loss}} = \eta_{\mathrm{t}} \boldsymbol{V}_0 + (1 - \eta_{\mathrm{t}}) \boldsymbol{I}_{2m}, \tag{14}$$

where $\boldsymbol{V}_0$ corresponds to the output state in the ideal case. It is believed that $\boldsymbol{V}_{\mathrm{loss}}$ can be decomposed into a classical part and a quantum part with lower entanglements, which may be efficiently approximated in a degree with methods such as the matrix product state [25, 48]. As a result, lossy GBS will become easier to simulate with the decrease of $\eta_{\mathrm{t}}$.

We first simulate small-scale lossy GBS, where we can compare the simulations with theories to ensure that our simulating programs are credible，as well as fully analyze the intrinsic data structures. In the ideal case, we can list all the output probabilities because the Hilbert space is well limited with a proper cutoff photon number $n_{\mathrm{cutoff}}$ in each mode [5]. As for the lossy case, however, direct calculations will become difficult because of mixed states, where we must calculate the hafnian of those matrices whose dimension is $2N$, instead of $N$ for those pure states. Here, we alternatively introduce the probability calculation method of balanced lossy GBS based on those probabilities of the ideal. This is feasible considering the fact that in lossy GBS, a certain output pattern may come from an ideal one with more photons, that is,

$$\mathrm{pr}_{\mathrm{loss}}(\boldsymbol{s}) = \sum_{\boldsymbol{s}'} \{\mathrm{pr}_{\mathrm{ideal}}(\boldsymbol{s}') \prod_{i=1}^{m} [\binom{s_i'}{s_i} \eta_{\mathrm{t}}^{s_i} (1 - \eta_{\mathrm{t}})^{s_i' - s_i}]\}. \tag{15}$$

$\mathrm{pr}_{\mathrm{ideal}}(\cdot)$ denotes the ideal output probability without loss, which can be obtained by Eq. (6) and (7). $\boldsymbol{s}'$ runs through all the patterns with $s_i' \geq s_i$ for all the modes. In each mode, the photon number $s_i$ conforms to an individual binomial distribution. Indeed, Eq. (15) has excluded the scenario that the pattern is degenerated from those with photons more than the cutoff number. However, this is safe because the corresponding probability is exponentially low. The small-scale numeric results (shown in the supplementary materials) indicate that our simulations are in agreement with theories.

Since our simulations have shown compatibilities with theories for the small-scale lossy GBS case, we can extend the pattern recognition validation on these simulating samples, to provide a quantificational relationship between the validation result and the loss level. In the validation approach, clusters should be constructed through sample training based on bona fide samples such as those from the ideal distribution. Among various cluster constructing methods, K-means++ is remarkable because of the learning ability. With K-means++, $k$ clusters are formed where $k$ is carefully adjusted to make the clusters characteristic and stable. The cluster quality can be evaluated through counting the bona fide sample number in each cluster. Here, for a robust cluster structure, the sample number distribution should be very unbalanced. Note that the pattern recognition is an approach computationally efficient but sample inefficient because of the dependence on the Hilbert space dimension. Therefore, in order to get a good validation performance, both the quantities of bona fide samples and samples to be tested should be sufficient. Here, a sample-box strategy [39] has been employed to partly improve sample efficiencies, where samples are put back into a box for reusing, once a test is over. Generally, a series of Gaussian peaks will be formed through counting the test value $\chi^2$, which can be expressed as [38]

$$\chi^2 = \sum_{i=1}^{k}\sum_{j=1}^{2}[(N_{ij} - E_{ij})^2/E_{ij}], \tag{16}$$

with $E_{ij} = N_i N_j/k$. $N_i = \sum_{j=1}^{2} N_{ij}$ denotes the sample number in the $i$th cluster and $N_j = \sum_{i=1}^{k} N_{ij}$ denotes the total number of the bona fide samples or the samples to be tested, corresponding to $j = 1$ or 2, respectively. At the beginning, we keep numbers of the two kind samples equal, whereas sample numbers in clusters usually conform to $N_{j=1} \geq N_{j=2}$ because some samples may be abandoned during sample sending. Before discussing the photon loss cases, we perform validations on ideal GBS using squeezed states as the input and some other mockups such as those using thermal, coherent [36] or squashed states [32] as the input. All the samples are obtained by classical simulation algorithms such as the rejection [49] and the chain rule. The Gaussian peaks are quite different in these cases. (For details, please see the supplementary materials.)

The extended validation results for photon loss cases with $\eta_{\mathrm{t}} \geq 0.9$ are presented in Fig. 1 (a). Despite the loss rate is kept low, it is well correlated with the Gaussian peak center $X_c$, forming an almost-linear relationship. The growing tendency can be explained by the facts that the distribution of bona fide samples in clusters is very unbalanced and that samples to be tested not belong to any cluster may become more with the loss rate [39]. Based on the relationship, once we perform a validation on lossy GBS where we only know the basic information such as the input SMSS number $K$, the squeezing parameter $r$ and the interferometer matrix $\boldsymbol{T}$, we can know the loss rate immediately. Note that the loss rate can also be evaluated roughly by simply counting output photon numbers (shown in the supplementary materials), whereas it is not so stable without learning output details.

In fact, the almost-linear relationship is a reflection of the intrinsic data structures of outputs. Owing to the small scale, we can list all the output probabilities with different $\eta_{\mathrm{t}}$ through Eq. (15). We can further evaluate the total of those high probabilities, for example, the total of the top 10. As shown in Fig. 1 (b), with increasing $\eta_{\mathrm{t}}$, the total probability is also increased. This indicates that photon loss will suppress the unbalance of GBS outputs in terms of probability distributions, where the distribution is very unbalanced for the ideal case. Meanwhile, the mean 2-norm distances $\overline{L_2}$ can also be obtained with $\overline{L_2} = \sum_{i=1}^{N_{\mathrm{out}}}[L_2\,(\boldsymbol{s}_i, \boldsymbol{s}_1)\mathrm{pr}\,(\boldsymbol{s}_i)]$, where the output patterns are arranged according to probabilities from high to low. $N_{\mathrm{out}} = (n_{\mathrm{cutoff}} + 1)^m - 1$ or $N_{\mathrm{out}} = \lfloor (n_{\mathrm{cutoff}} + 1)^m/2 - 1/2 \rfloor$ is the total output pattern number in the Hilbert space with or without loss, where the zero states have been excluded. $L_2(\boldsymbol{s}_i, \boldsymbol{s}_1) = [\sum_{j=1}^{m}(s_{ij} - s_{1j})^2]^{1/2}$ denotes the 2-norm distance between the $i$th pattern $\boldsymbol{s}_i$ and the pattern $\boldsymbol{s}_1$ of which the probability is the highest. It is found that $\overline{L_2}$ also changes almost-linearly with $\eta_{\mathrm{t}}$, as shown in Fig. 1 (c). More specially, if we look at those patterns which have a much shorter or a much longer norm distance from $\boldsymbol{s}_1$, we can find that the corresponding probabilities will become higher with $\eta_{\mathrm{t}}$. Hence, photon loss will also suppress the unbalance in terms of norm distances, where those patterns with obviously shorter or longer norm distances tend to become dominant when the loss rate is low. When the scale is larger, the extended validation approach is still useful to lossy GBS, as shown in Fig. 1 (d), whereas obviously more samples and correspondingly a much larger cluster number are in need because of sample inefficiencies. Furthermore, the cases with photon loss happening in optical units are also discussed in the supplementary materials, where the Gaussian peak center monotonicity with loss rates still holds.

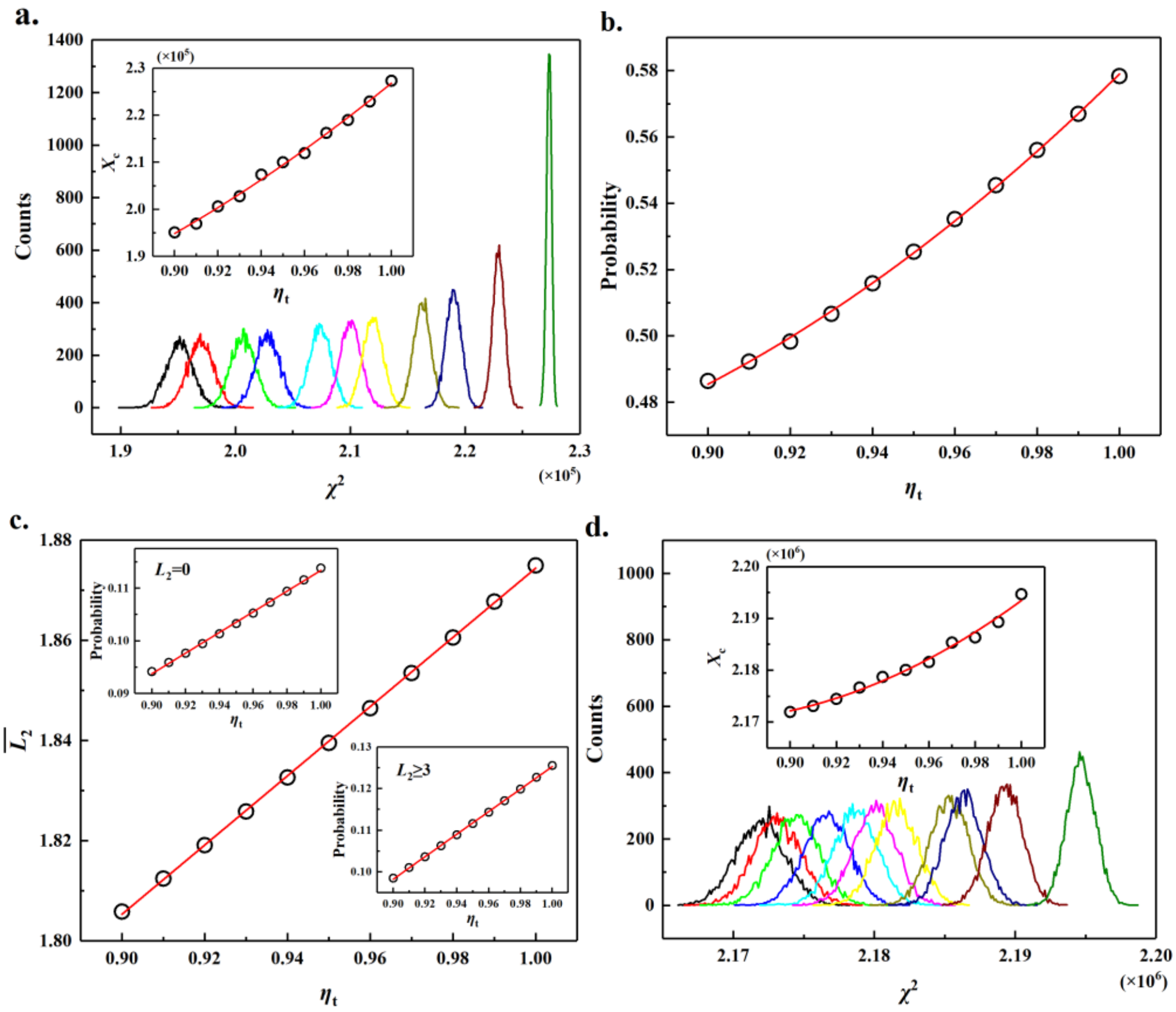

Fig. 1 Extended pattern recognition validations on simulated lossy GBS and output data structure characterizations. (a) Gaussian peaks formed by repeating calculating $\chi^2$ for $10^4$ times using a sample-box strategy, in small-scale GBS cases with $r = 0.5$, $K = m = 5$, $n_{\text{cutoff}} = 4$ and the total sample number is $10^4$ for each $\eta_{\text{t}}$. With increasing $\eta_{\text{t}}$, the Gaussian peak shifts to the right. As for the validation parameters, $k$ is set as 150 and the training sample number is 3000, with the ideal samples used as the bona fide ones. The inset shows the almost-linear relationship of the Gaussian peak center $X_{\text{c}}$ and $\eta_{\text{t}}$. (b) Plots of the total of the top 10 high output probabilities and $\eta_{\text{t}}$, in small-scale cases. (c) Corresponding plots of the mean 2-norm distance $\overline{L_2}$ and $\eta_{\text{t}}$. The insets show total probabilities of those patterns with $L_2 = 0$ and with $L_2 \geq 3$, respectively. (d) Validation results in larger-scale cases where $r = 0.2$, $K = m = 10$, $n_{\text{cutoff}} = 3$ and the total sample number is $10^5$. To get a good validation performance, $k$ is increased to 700 and the training sample number is increased to $2\times10^4$. The red lines in those relevant graphs are fitted curves.

# 5 Extended validations on GBS with photon distinguishability

Photon distinguishability is another main kind of noises which can also be employed to realize classical simulating speedup. In a typical model, when those partially distinguishable photons enter into the interferometer, the process can be equally described as indistinguishable photons entering into actual modes along with some totally distinguishable photons entering into virtual modes [22]. The output is a combination of these results, that is, one term from the actual modes and *K* terms from the virtual modes. For the actual part, it is similar to a photon loss model which may be simplified as mentioned in Sec. 4. For the virtual part, it is found that when a certain mode has been chosen as the SMSS input mode, the output photons are related to an individual multinomial distribution. Therefore, the virtual part is totally classical.

Let us go back to the exact simulation, where one process of the actual part and *K* processes of the virtual part are carried out simultaneously. In the ordering of $(\hat{\boldsymbol{x}}, \hat{\boldsymbol{p}})$, the covariance matrices of input states corresponding to the two parts can be written as [22]

$$\boldsymbol{V}_{\text{actual}}^{(0)} = \eta_{\text{ind}}\boldsymbol{V}_{\text{in}} + (1-\eta_{\text{ind}})\boldsymbol{I}_{2m}, \tag{17}$$

$$\boldsymbol{V}_{\text{virtual}}^{(i)} = (1-\eta_{\text{ind}})[(\boldsymbol{I}_{i-1} \oplus e^{2r_i} \oplus \boldsymbol{I}_{m-i}) \oplus (\boldsymbol{I}_{i-1} \oplus e^{-2r_i} \oplus \boldsymbol{I}_{m-i})] + \eta_{\text{ind}}\boldsymbol{I}_{2m}, \tag{18}$$

where $i = 1, \dots, K$. In Eq. (17), $\boldsymbol{V}_{\text{in}}$ corresponds to the ideal case which can be expressed via Eq. (2). $\eta_{\text{ind}}$ denotes the probability of photons remaining indistinguishable, and thus, reveals the noise level of photon distinguishability. Once those photons are transformed to be totally distinguishable, as described by Eq. (18), they will be sent into the *i*th virtual mode, where *i* is the original input mode number.

The output pattern is contributed by one pattern from actual modes and *K* patterns from virtual modes. Therefore, the output probability with partial distinguishability is [22]

$$\text{pr}_{\text{pd}}(\boldsymbol{s}) = \sum_{\boldsymbol{s}^{(0)},\dots,\boldsymbol{s}^{(K)}}[\text{pr}_{\text{actual}}(\boldsymbol{s}^{(0)})\prod_{i=1}^{K}\text{pr}_{\text{virtual}}(\boldsymbol{s}^{(i)})], \tag{19}$$

where $\boldsymbol{s}^{(0)} + \dots + \boldsymbol{s}^{(K)} = \boldsymbol{s}$. $\text{pr}_{\text{actual}}(\cdot)$ and $\text{pr}_{\text{virtual}}(\cdot)$ denote the output probabilities with input states presented by Eq. (17) and (18), respectively. In practice, we list probabilities of all the output patterns from $(0,..,0)$ to $(n_{\text{cutoff}}, \dots, n_{\text{cutoff}})$ for the $K+1$ processes. Hence, there are $(n_{\text{cutoff}}+1)^{m(K+1)}$ patterns we need to consider to obtain the final probabilities. We further exclude those final probabilities corresponding to patterns with photons more than the cutoff number and with $N = 0$. After that, we make a normalization. These operations are time-consuming because the pattern number to be considered grows exponentially with both *m* and *K*. In order to ensure that our simulations are in agreement with theories, however, we can consider only the marginal probabilities of patterns with the maximum photon number in each mode $n_{\max} < n_{\text{cutoff}}$ (shown in the supplementary materials).

The extended validation results for photon distinguishability cases with $\eta_{\text{ind}} \geq 0.9$ are presented in Fig. 2 (a). The growing tendency of Gaussian peak centers with $\eta_{\text{ind}}$ is similar with that of photon loss, except that the growing trace is more curved. The raised tail hints that only with $\eta_{\text{ind}}$ obviously high, those high order multiphoton interferences may become significant, to which the pattern recognition validation approach is found to be sensitive. The validation results in high-noise regimes are also provided in the supplementary materials, where the overall tendencies are different more obviously for the two main kind noises.

We further investigate the intrinsic data structures where we keep *K* lower to realize all the pattern probability calculations based on Eq. (19). As shown in Fig. 2 (b), the total of those high probabilities also increases with $\eta_{\text{ind}}$, revealing the similar tendency with that of validation results. This indicates that the unbalance of GBS outputs may also be suppressed by photon distinguishability. The tendency similarity hints that the unbalance is a natural result of multiphoton interferences, in terms of probability distributions. More precisely, it is a result of destructive and constructive interferences. The mean norm distance, however, only looks like to have an almost-linear relationship with $\eta_{\text{ind}}$, as shown in Fig. 2 (c). In detail, we also provide statistics of those patterns whose 2-norm distance is obviously shorter or longer. Both grow gradually with $\eta_{\text{ind}}$, indicating the unbalance distribution in terms of norm distances can also be suppressed by photon distinguishability. Actually, the mean norm distance change is a competing result of those shorter and longer ones, only showing a monotonous tendency. We also extend validations in the larger-scale cases with photon distinguishability, as shown in Fig. 2 (d), where the parameter requirement is similar to that of the lossy GBS. In addition to the two main kind noises, the extended validations on GBS with the noise of dark counts are also discussed in the supplementary materials.

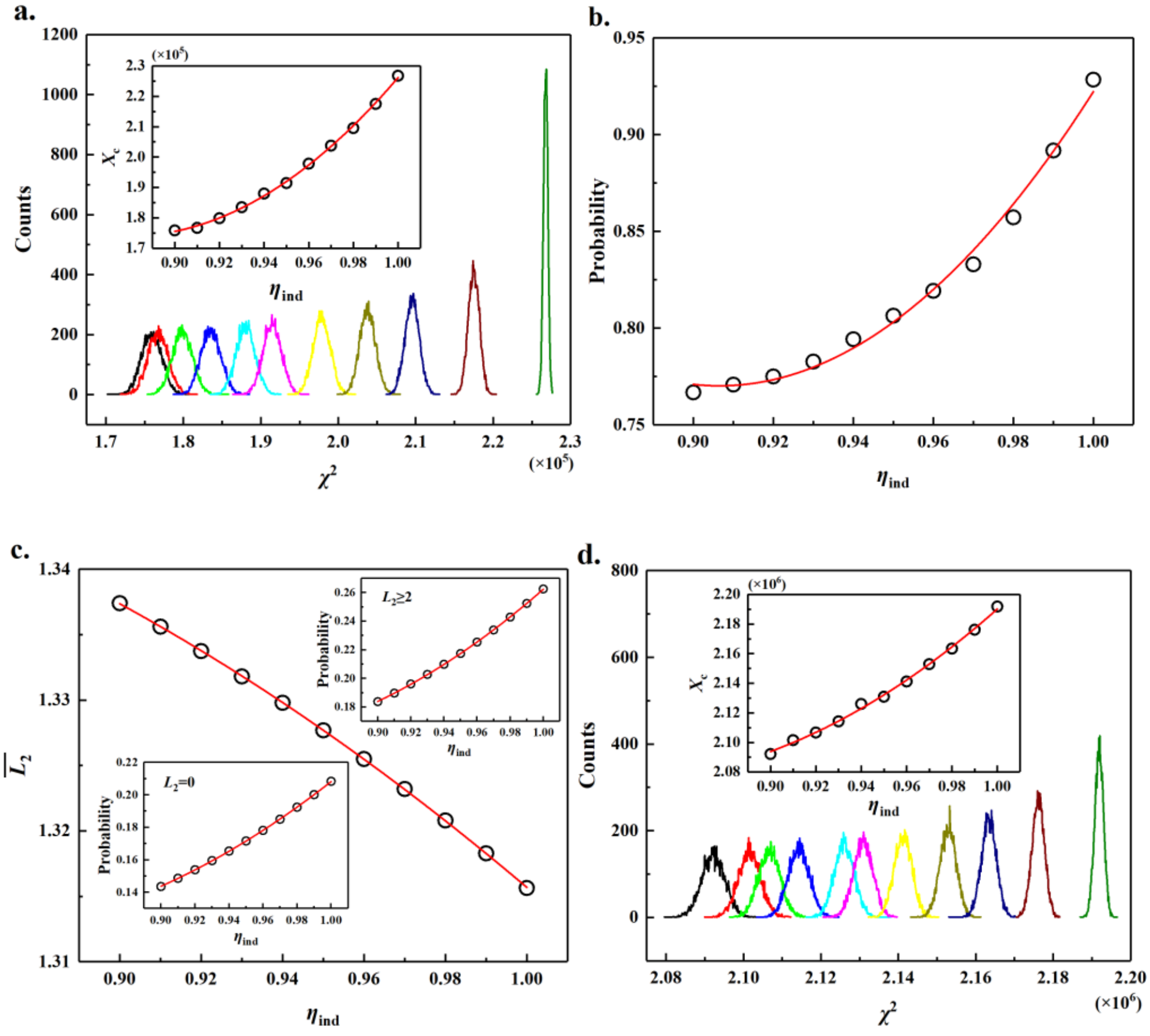

Fig. 2 Extended pattern recognition validations on simulated GBS with photon distinguishability and output data structure characterizations. (a) Validations in small-scale cases using sample and validation parameters the same as those in small-scale lossy cases. (b) Plots of the total of those high probabilities and $\eta_{\text{t}}$, in the cases where $r = 0.15,\ K = 3,\ m = 5$ and $n_{\text{cutoff}} = 2$. (c) Corresponding plots of the mean 2-norm distance $\overline{L_2}$ and $\eta_{\text{t}}$. The insets show total probabilities of those patterns with $L_2 = 0$ and with $L_2 \geq 2$, respectively. (d) Validation results in larger-scale cases using parameters the same as those in larger-scale lossy cases. The red lines in those relevant graphs are fitted curves.

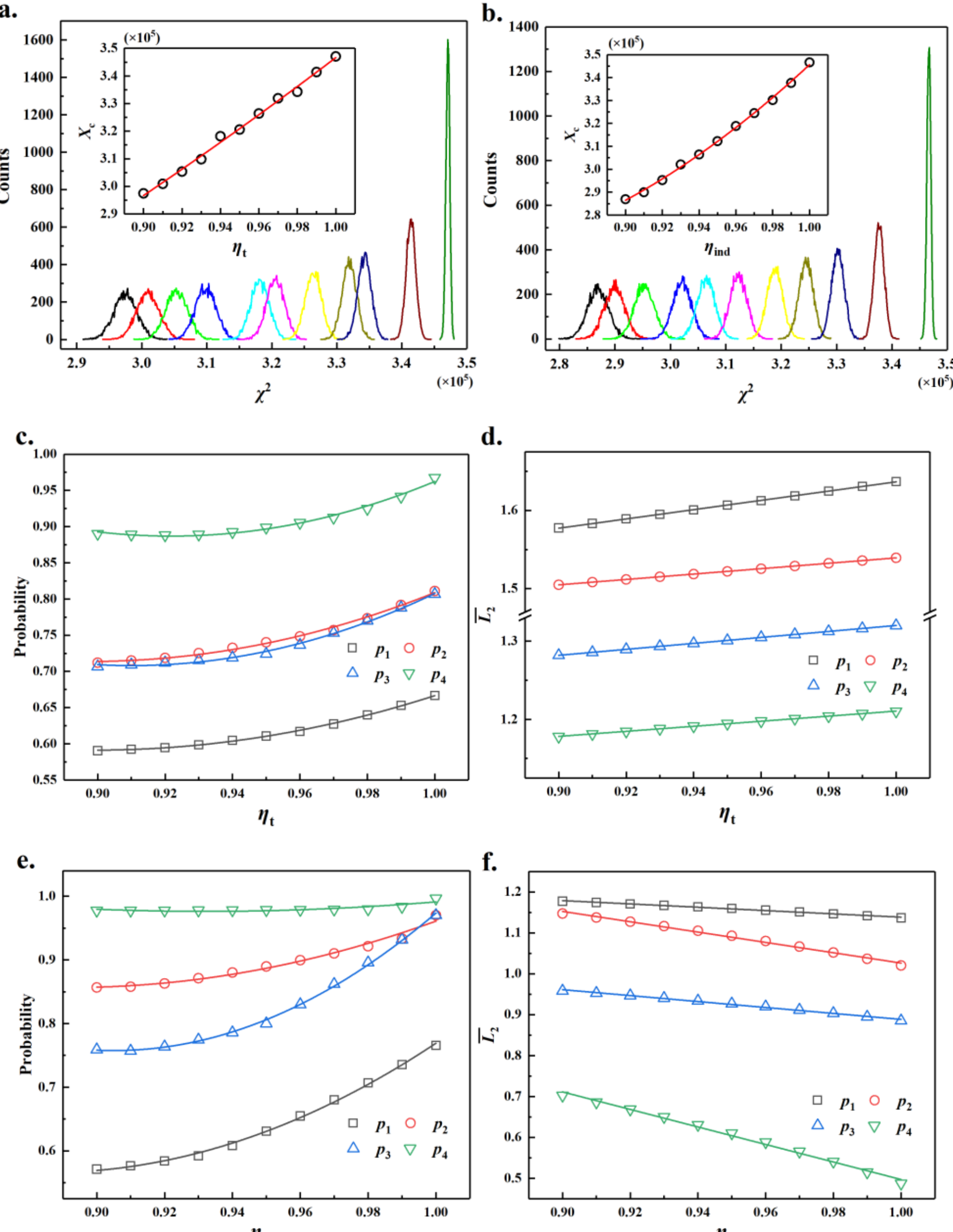

Fig. 3 Extended pattern recognition validations on larger-scale noisy GBS using the output-binning strategy and characterizations. (a) Validations in lossy cases. (b) Validations in distinguishability cases. In these cases, the adjacent modes are bound together as a subset. Hence the binned output mode combination is $\{\{1,2\},\ldots,\{9,10\}\}$. Owing to the strategy, the good-performance parameter requirement is greatly relaxed with $k = 100$ and the required total and training sample numbers are only $10^4$ and 3000. (c)-(f) Total of top 10 high probabilities and mean 2-norm distance changes of typical binned output mode partitions in small-scale cases, with $\eta_t$ and $\eta_{ind}$, respectively. The partitions are $p_1 = \{\{1,2\},\{3\},\{4\},\{5\}\}$, $p_2 = \{\{1,2,3\},\{4\},\{5\}\}$, $p_3 = \{\{1,2\},\{3,4\},\{5\}\}$ and $p_4 = \{\{1,2,3,4\},\{5\}\}$. The small-scale sample parameters are $r = 0.5$, $K = m = 5$, $n_{cutoff} = 4$ in lossy cases and $r = 0.15$, $K = 3$, $m = 5$, $n_{cutoff} = 2$ in distinguishability cases. The solid lines in relevant graphs are fitted curves.

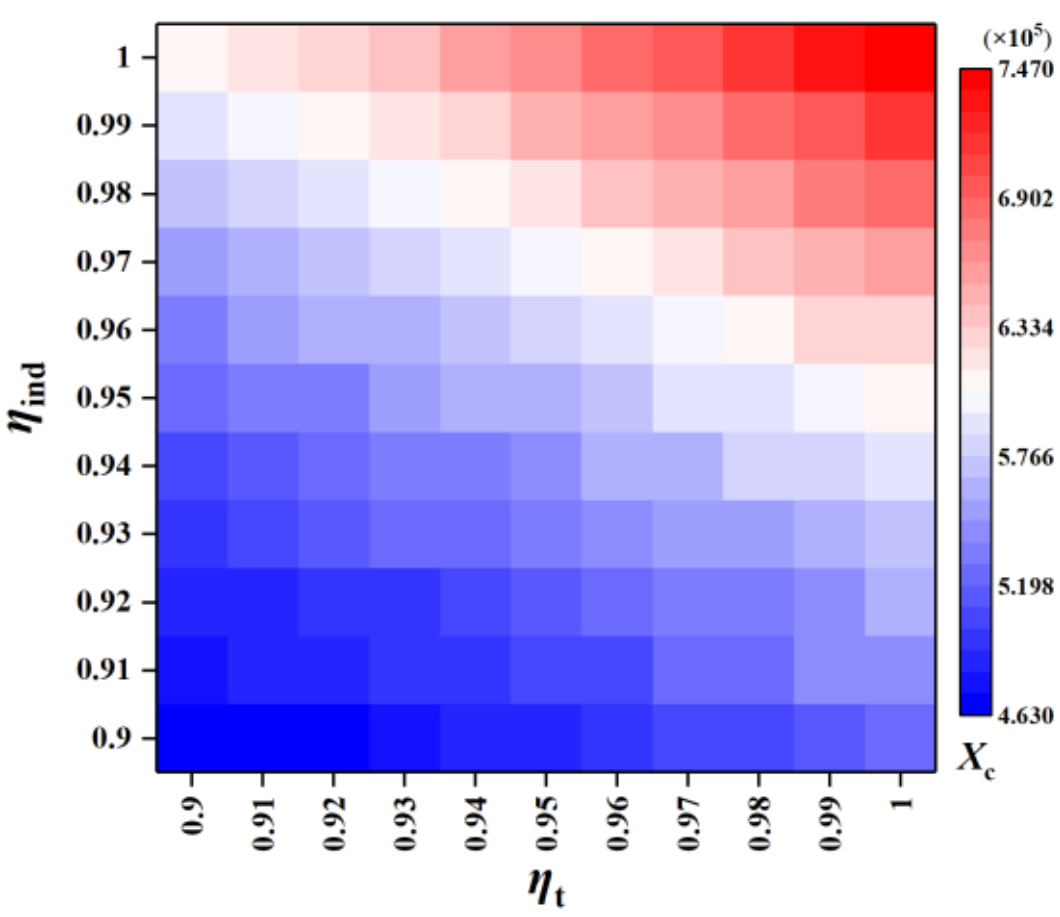

Fig. 4 The Gaussian peak center map with both of the main noises, where the output-binning strategy is employed. For each pure color sub-box, the number of test samples is 2×10$^4$, which are generated by randomly selecting from 10$^5$ exactly simulated samples and post treating. The basic sample parameters are kept the same as those in the larger-scale cases. The adjacent modes are bound together as those in the cases of Fig. 3. The cluster number $k$ is 130 and the training sample number is 5000.

# 6 The output-binning strategy

It has been reported that in the output-binning validation approach, photon number distributions of those binned outputs are found to be different for GBS with and without noises [35]. This motivates us to employ the output-binning strategy before performing the pattern recognition validations, where output photons in certain modes are bound together to be measured. As a result, the Hilbert space dimension of the output states is greatly reduced, which we can utilize to realize validation speedup. As for the noisy GBS with binned outputs represented by $l$ subsets ($l < m$), the Hilbert space dimension including the zero state is $\prod_{i=1}^{l}(m_{\mathrm{sub},i}n_{\mathrm{cutoff}}+1)$, where $m_{\mathrm{sub},i}$ is the mode number in the $i$th subset with $\sum_i m_{\mathrm{sub},i} = m$. This is obviously smaller than the original dimension $(n_{\mathrm{cutoff}}+1)^m = \prod_{i=1}^{l}(n_{\mathrm{cutoff}}+1)^{m_{\mathrm{sub},i}}$. Hence, the data structures become easier to characterize, where the cluster number $k$ and correspondingly the bona fide sample number can be greatly reduced, and thus the validation is faster, considering the fact that its computational complexity is polynomial to both $k$ and the sample number.

The output-binning strategy seems to be not so harmful to the resultant monotonicity in Fig. 3 (a) and (b). This is based on the fact that the distribution unbalances resulted from interferences still hold with counting statistics for those subsets, partly considering that the unbalanced total photon number distribution is fixed. Therefore, it is safe to employ the output-binning strategy properly without destroying the unbalances. We further provide numeric statistics of high probabilities and mean norm distances in those small-scale cases, where typical partitions of the mode number list $\{1, \ldots, m\}$ are considered. As shown in Fig. 3 (c)-(f), the monotonicity with noise levels of both photon loss and photon distinguishability, still exists in the intrinsic output data structures, on the two aforementioned aspects.

Based on the strategy, we furtherly provide the validation results for weakly-correlated larger-scale GBS samples simultaneously influenced by both of the main noises. The samples are approximated, generated by randomly selecting from those exactly simulated samples with the noise of photon distinguishability, and then introducing photon loss according to the binominal distributions with Eq. (15). Owing to the robust clusters based on K-means++, which well characterize the output data structures, the validation performance is still stable, as shown in Fig. 4, where all the values of $X_c$ follow the overall tendencies, in both horizontal and vertical directions, even with much fewer samples compared to those larger-scale cases without employing the output-binning strategy.

With the strategy, the extended validation approach can deal with large-scale samples efficiently, especially on the condition where only the mode number is increased. However, when the photon number is also increased, the approach may be invalid because of the lack of classically intractable ideal samples to act as the bona fide ones. However, those approximation methods such as the matrix product state [25, 48] may be employed to generate samples for cluster constructing. These approximate samples must be different from those physically noisy counterparts. (For details, please see the supplementary materials.)

# 7 The extended correlation validation as a comparison

Last but not least, we perform an extension of the correlation validation approach [31-33], which is also computationally efficient, as a comparison. The sample numbers we firstly used are the same with the corresponding normal pattern recognition validations. Hence, at least in this paper, we do not explore any evidences that the correlation approach may be sample efficient, or the output-binning strategy is suitable to this approach in other words. As for the *t*-order correlation, *t* output modes $(o_1, \dots, o_t)$ are chosen beforehand without repetitions to generate a point. Hence, there are $\binom{m}{t}$ points in total. With respect to the two kind samples for comparisons, the point coordinates can be obtained by calculating the correlation function [32]

$$\kappa\left(n_{o_1}, \dots, n_{o_t}\right) = \sum_\pi [(|\pi| - 1)!\,(-1)^{|\pi|-1} \prod_{B\in\pi} \langle \prod_{i\in B} n_{o_i} \rangle]. \tag{20}$$

$\pi$ runs through all the partitions of $\{1, \dots, t\}$, $n_{o_i}$ is the photon number in the mode $o_i$. When $m$ is high, $t$ should be kept low to realize efficient computations. Here, the validations are carried out with $t$ up to 4. Instead of looking at those correlation coefficients which may only be useful when mockups are very different, we relate the validation results with noise levels through evaluating the coordinate deviations, simply associated with $\Gamma = (\sum_i |\kappa_{\mathrm{noise},i}|)/(\sum_i |\kappa_{\mathrm{ideal},i}|)$. $\kappa_{\mathrm{noise},i}$ and $\kappa_{\mathrm{ideal},i}$ are the correlation coordinates of the *i*th point. Note that regular deviations are very inapparent but do exist when noises are low. To make it clear, we present validation results in Fig. 5, with noise levels kept obviously different. Based on the correlation approach, both of the noises are evaluated through the monotonous changes of $\Gamma$, whereas higher order correlators are more sensitive to noises in Fig. 6, which are actually reflections of multiphoton interference degrees. However, in general, those higher order correlators need much more calculations, considering $\binom{m}{t}$ terms in addition to the computational complexity according to the definition of Eq. (20).

If the sample numbers are deliberately decreased, however, it is found that the pattern recognition validation with output-binning strategy can numerically outperform the normal correlation validation, where both of the main noises are simultaneously considered (shown in the supplementary materials). This comparison indicates that the former may be more sample efficient.

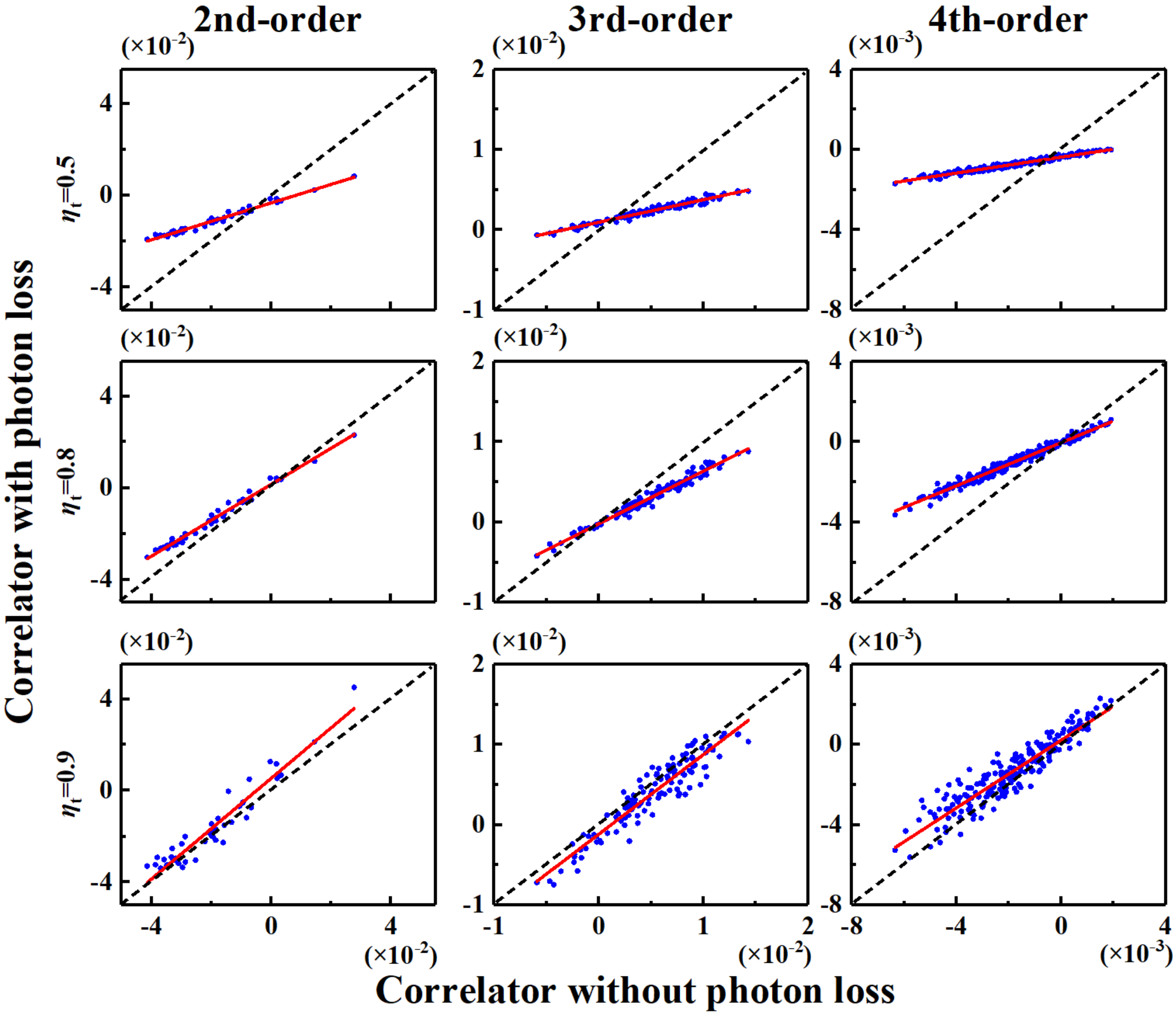

Fig. 5 Correlators for larger-scale lossy GBS where $\eta_t$ values are very different. The sample parameters such as the total sample number are kept the same with those employed in original pattern recognition validations. The red solid lines are fitted lines according to the points.

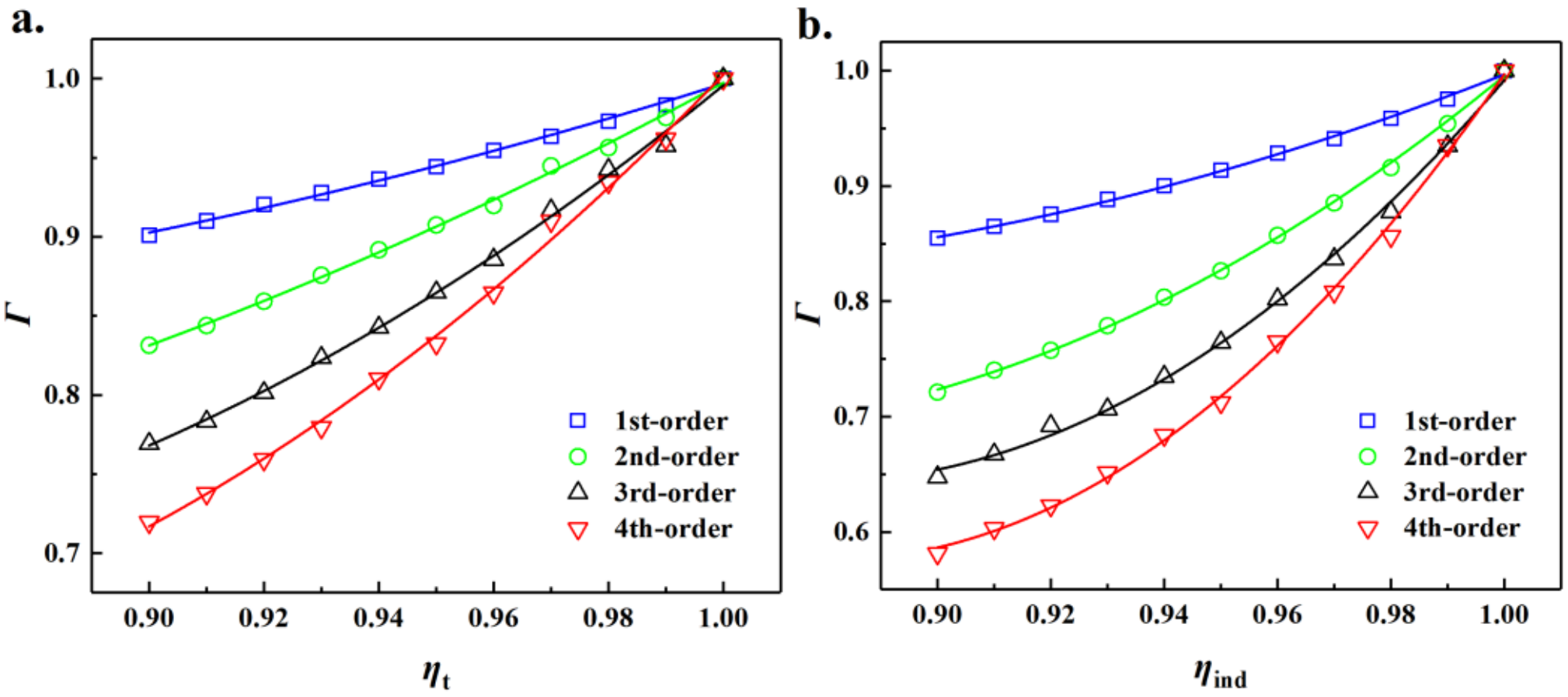

Fig. 6 Plots of coordinate deviations evaluated by $\Gamma$ and noise levels in larger-scale cases, based on 1st- to 4th-order correlators. (a) Cases with photon loss. (b) Cases with photon distinguishability. The solid lines are fitted curves.

# 8 Conclusion

In conclusion, we extend validations on exactly simulated PNRD based GBS with main noises of photon loss and photon distinguishability, using pattern recognition techniques. Owing to the advances such as thorough characterizations of intrinsic data structures, including the unbalances resulted from multiphoton interferences, the extended approach provides a quantificational relationship between validation results and noise levels, even when noises are sufficiently low. We employ an output-binning strategy in order to overcome the drawback of sample inefficiencies. We also extend the correlation approach as a comparison. Our simulations indicate that pattern recognition validation approach is sensitive to noises and it may be helpful to evaluate noises in some other quantum computation models, where finite noises are dominant.

# Acknowledgements

This work can be carried out owing to Project supported by Shanghai Municipal Science and Technology Major Project (Grant No. 2019SHZDZX01).

# Supplementary materials for "Extended validations on photon number resolving detector based Gaussian boson sampling with low noises"

Yang Ji[1,2], Yongzheng Wu[1,2], Shi Wang[1,2], Jie Hou[1,2], Zijian Wang[1], Bo Jiang[*1]
[1]*The 32nd Research Institute of China Electronics Technology Group Corporation, Shanghai 201808, China*
[2]*Shanghai Research Center for Quantum Sciences, Shanghai 201315, China*
*Corresponding author. E-mail: b26jiang@126.com

## 1 Simulation evaluations

In order to ensure that our exact simulations of noisy Gaussian boson sampling (GBS) are reliable, we make comparisons of simulations and theories. As for GBS with balanced photon loss, the probability of output patterns can be obtained via Eq. (15) in the manuscript. We simulate a small-scale case with loss, where we count the samples to form a distribution. As shown in Fig. S1 (a), the simulations are in agreement with theories.

On the other hand, for GBS with photon distinguishability, obtaining all the probabilities is hard when Eq. (19) in the manuscript is used. We only calculate those marginal probabilities with the maximum photon number in each mode $n_{\max} < n_{\text{cutoff}}$. Therefore, only $(n_{\max} + 1)^{m(K+1)}$ patterns should be considered. The marginal probability scheme is suitable to those photon distinguishability cases with $r$ kept relatively low, where output patterns with small photon numbers become dominant in those simulated samples and thus the corresponding marginal probabilities are more reliable, when the total simulated sample number is finite. The simulations are still in agreement with theories, as shown in Fig. S1 (b).

As a supplement, we obtain the mean output photon numbers of lossy GBS, keeping optical parameters and sample numbers the same as those employed in Fig. 1 (a) in the manuscript. As shown in Fig. S1 (c), the overall trend is monotonous but not so smooth as that in Fig. 1 (a) and some points may not follow the trend. This is because the sample number is still not sufficiently large.

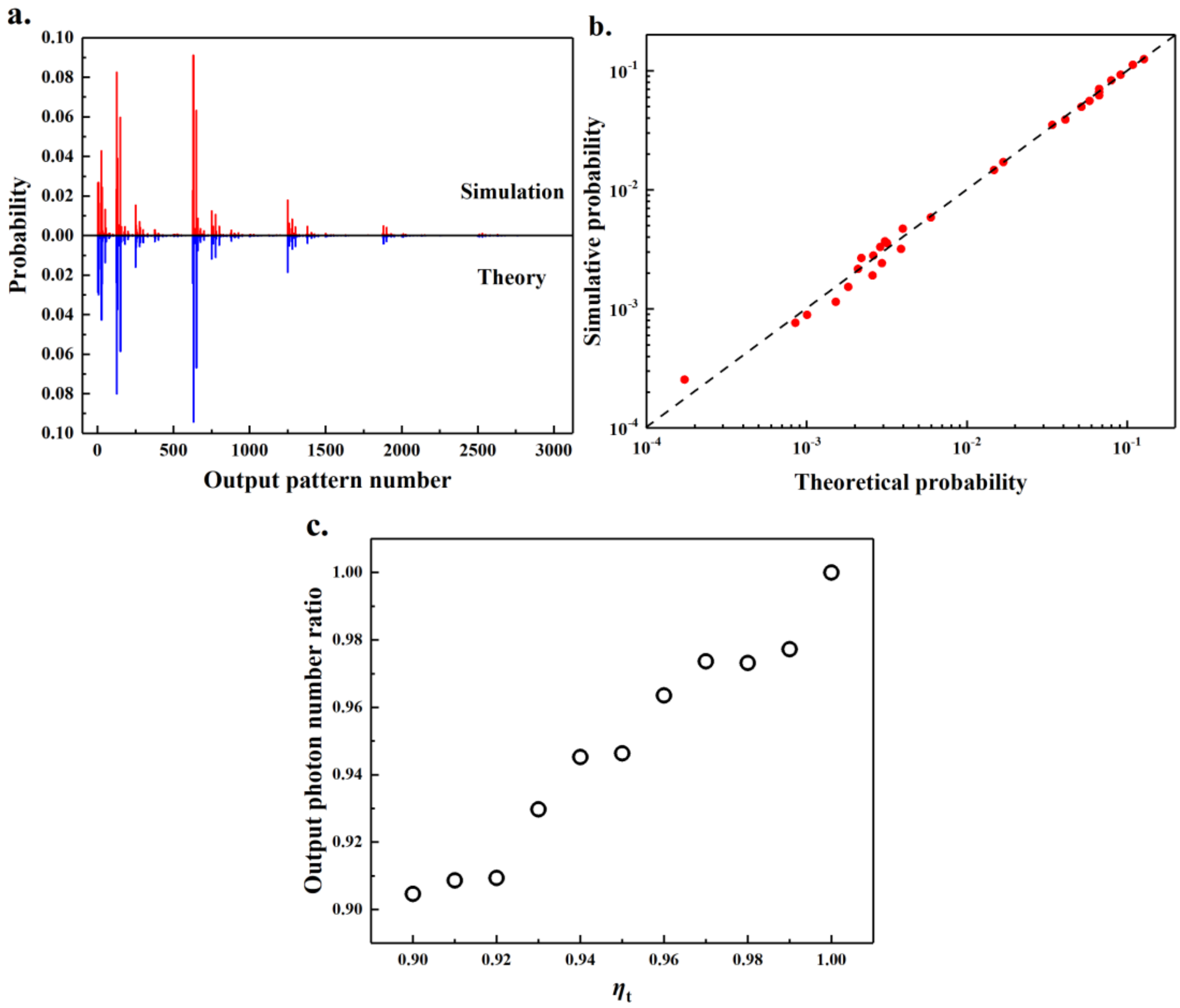

Fig. S1 Evaluations of the simulations of small-scale noisy GBS. (a) Output probabilities obtained by simulations and theoretical calculations for lossy GBS, where $r = 0.5$ and $\eta_t = 0.9$. (b) Marginal output probabilities of GBS with photon distinguishability, with the maximum photon number $n_{\max} = 1$ in each mode, where $r = 0.3$ and $\eta_{\text{ind}} = 0.9$. For each kind of noises, the simulated sample number is $10^4$. (c) Plots of output photon number ratio and $\eta_t$, with sample parameters the same as those used in Fig. 1 (a) in the manuscript, where for each $\eta_t$, the sample number is also $10^4$.

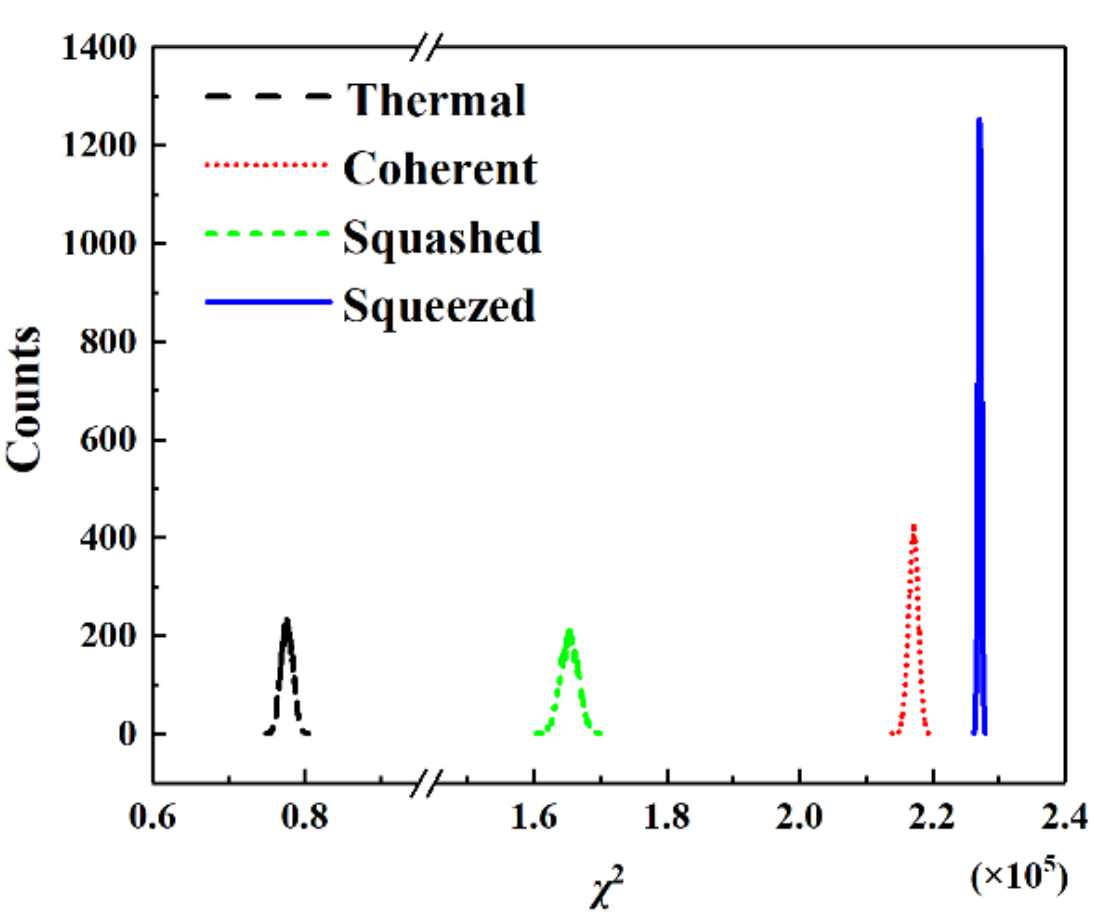

Fig. S2 Pattern recognition validations on GBS using the squeezed states as the input and other mockups using thermal, coherent and squashed states as the input. The ideal GBS sample parameters and the validation parameters are the same as those used in Fig. 1 (a) and Fig. 2 (a) in the manuscript.

# 2 Validations on GBS and other mockups

Before extending validations on noisy GBS, we preform the pattern recognition validations to discriminate ideal GBS using squeezed states as the input and some other mockups using thermal, coherent or squashed states as the input.

The probability of an output pattern $\boldsymbol{s}$ for the thermal-state input mockup is [1]

$$\mathrm{pr}_{\mathrm{th}}(\boldsymbol{s}) = \frac{\mathrm{perm}(\boldsymbol{D}_s)}{\prod_i s_i! \prod_i (1+\langle n_i \rangle)}, \tag{S1}$$

where $\boldsymbol{D} = \boldsymbol{T}\{\oplus_i [\langle n_i \rangle / (1 + \langle n_i \rangle)]\}\boldsymbol{T}^{\dagger}$, $\langle n_i \rangle$ is the mean photon number in the $i$th mode, which is set as [2]

$$\langle n_i \rangle = \sinh^2 r_i = \sinh^2 r. \tag{S2}$$

This means that the mean photon number in each input mode is kept equal for GBS and the compared mockups. $\boldsymbol{D}_s$ is a submatrix of $\boldsymbol{D}$ constructed like $\boldsymbol{B}_s$ in the manuscript.

The probability for the coherent-sate input mockup is [1]

$$\mathrm{pr}_{\mathrm{co}}(\boldsymbol{s}) = \prod_i \frac{e^{-|\beta_i|^2} |\beta_i|^{2s_i}}{s_i!}, \tag{S3}$$

where $\beta_i = \sum_j T_{ji} \alpha_j$. $\alpha_j$ represents the coherent state in the $j$th input mode. Generally, it can be written as

$$\alpha_j = \sqrt{\langle n_j \rangle} e^{i\theta}. \tag{S4}$$

Here we set $\theta = 0$. With these probabilities, samples can be generated for the two kind small-scale mockups, using the rejection algorithm [3].

As for the mockup using squashed states as the input, where one canonical term only has the vacuum fluctuations and the other term has fluctuations related to the mean photon number, the input covariance matrix is [4]

$$\boldsymbol{V}_{\mathrm{sq}} = [\oplus_{i=1}^{K} (1 + 4\langle n_i \rangle)] \oplus \boldsymbol{I}_{2m-K}. \tag{S5}$$

With this, we can use the chain rule method to generate samples [5, 6]. We then perform the pattern recognition validations [7]. As shown in Fig. S2, the Gaussian peaks are quite different for GBS and the compared mockups, indicating that the validation approach is effective.

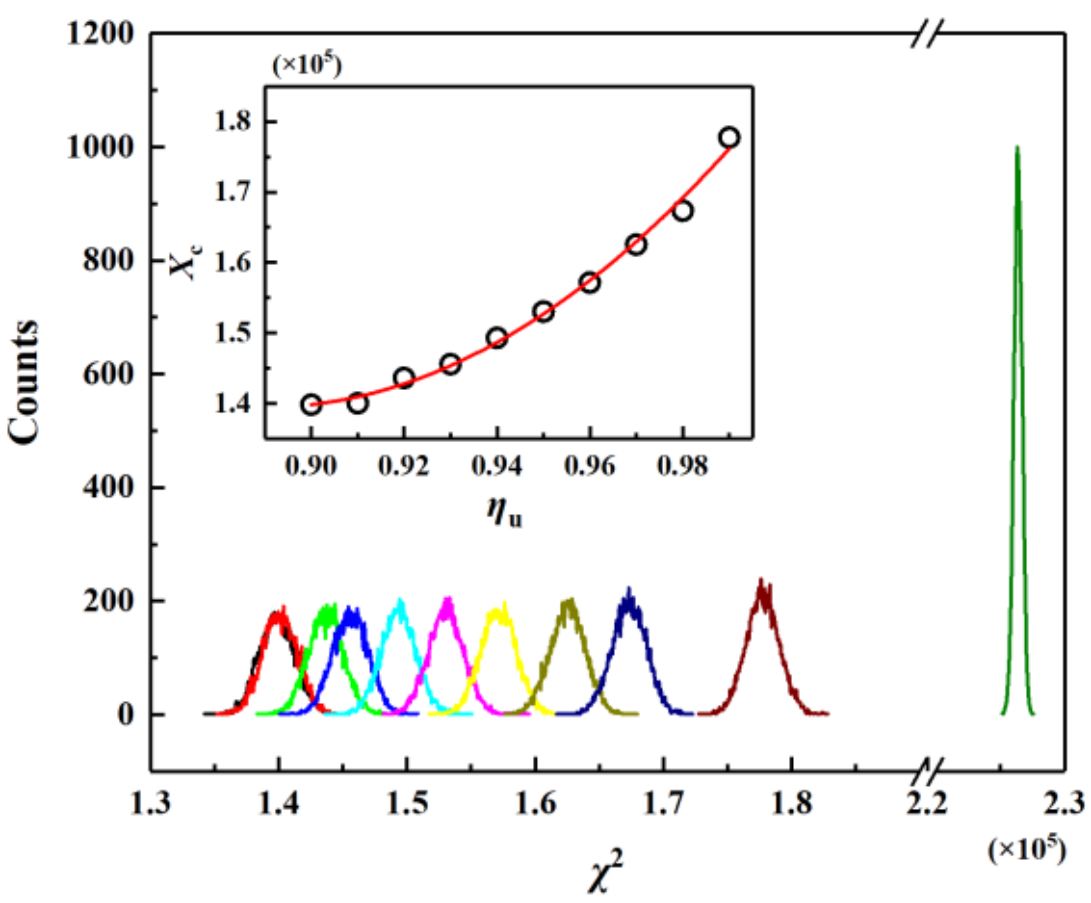

Fig. S3 Extended pattern recognition validations on GBS with photon loss happening in optical units. The basic sample and validation parameters are kept the same as those in small-scale cases in the manuscript.

# 3 Photon loss in optical units

We consider a perhaps more realistic experimental condition, by assigning the photon loss rate $1-\eta_{\mathrm{u}}$ to each optical unit in the interferometer [8], that is, the combination of a beam splitter and a phase shifter. $\eta_{\mathrm{u}}$ is the transmission rate of the unit. A symmetrical network is employed to be the interferometer [9]. Suppose the mode number is $m$, the $m$-dimensional random matrix $\boldsymbol{T}$ corresponding to the interferometer can be decomposed by

$$\boldsymbol{T} = \boldsymbol{D}\boldsymbol{M}_1\boldsymbol{M}_2 \dots \boldsymbol{M}_{m(m-1)/2}, \tag{S6}$$

where $\boldsymbol{D}$ is a diagonal matrix charactering the terminal phase shifters and $\boldsymbol{M}_i$ $(i = 1, \dots, m(m-1)/2)$ is the matrix charactering the $i$th optical unit, where two certain paths are connected. By noticing that once a photon is lost in the unit, it is equivalent to the fact that the photon enters into an environmental path through a virtual beam splitter, we can reconstruct the unit matrix as [8]

$$\boldsymbol{M}_i' = (\boldsymbol{M}_i \oplus \boldsymbol{I}_{m(m-1)})\boldsymbol{B}_{i1}\boldsymbol{B}_{i2}, \tag{S7}$$

where $\boldsymbol{I}_{m(m-1)}$ is the $m(m-1)$-demensional identity matrix, $\boldsymbol{B}_{i1}$ and $\boldsymbol{B}_{i2}$ are matrices of the two virtual beam splitters connecting one realistic path and one virtual path, of which the reflectivity is $\sqrt{\eta_{\mathrm{u}}}$. Hence, we can write down the overall interferometer matrix with loss taken into consideration as

$$\boldsymbol{T}' = (\boldsymbol{D} \oplus \boldsymbol{I}_{m(m-1)})\boldsymbol{M}_1{}'\boldsymbol{M}_2{}' \dots \boldsymbol{M}_{m(m-1)/2}{}'. \tag{S8}$$

The left work we need to realize the perhaps more realistic model is to expand input modes with those using vacuum states as the input and ignore those outputs occurring in virtual modes. The validation results are shown in Fig. S3, where the Gaussian peak center monotonicity with noises still holds. The peaks corresponding to noisy samples are relatively far away from that of the ideal, because of higher noises compared to the counterparts in balanced loss cases, since there are $m(m-1)/2$ units to consider.

# 4 Validation results in high-noise regimes

We further show the extended validation results, including those in high-noise regimes, for small-scale noisy GBS mentioned in the manuscript, with considering both photon loss and photon distinguishability. It is found that the validation results are quite different for the two kind noises, as shown in Fig. S4. The almost-linear relationship still holds with balanced photon loss while with photon distinguishability, Gaussian peak center location changes are only obvious when the probability of photons remaining indistinguishable $\eta_{\mathrm{ind}}$ is large enough. This tendency is similar to that shown in our previous work where validations are extended on boson sampling with photon distinguishability [10]. With smaller $\eta_{\mathrm{ind}}$, the Gaussian peaks are crowded and far away from that corresponding to the ideal samples, because of negligible contributions from quantum processes. This result hints once again that those experiments may be meaningless even if they can work with $\eta_{\mathrm{ind}}$ not too small but still far away from 1.

# 5 Dark counts

We only assume perfect photon number resolving detectors in the manuscript, because realistic probabilities of detectors being imperfect are exponentially small. For example, the dark count probability is below $10^{-6}$ [11], in which case detectors may count photons by mistake. Nevertheless, here we show two examples where dark counts are considered.

The first example is the extended validation result of GBS with only the noise of dark counts. Those noisy samples are generated by first simulating ideal samples and then adding photons according to dark count probabilities in each mode. Please note that different from those in reality, the dark count probabilities considered here are high enough to make dark counts significant in GBS. As shown in Fig. S5 (a), with decreasing $p_{\mathrm{dc}}$, the dark count probability in each mode, the Gaussian peak center location grows correspondingly. This example indicates that the extended pattern recognition validation is still useful to dealing with some other noises such as dark counts.

The second example is the extended validation result of GBS with noises of both photon loss and dark counts. We take care of this case because intuitively the total output photon numbers may do not fluctuate so dramatically with noises as those in single-noise cases, considering the fact that lost photons may be compensated by dark counts. Hence, it may be harder to evaluate noise levels. Nevertheless, the extended validation approach still performs well, as shown in Fig. S5 (b). This is perhaps because the unbalance of data structures may vanish with noises, considering the fact that the unbalanced structure is first destroyed due to photon loss, and then replaced by a more balanced one due to dark counts [12].

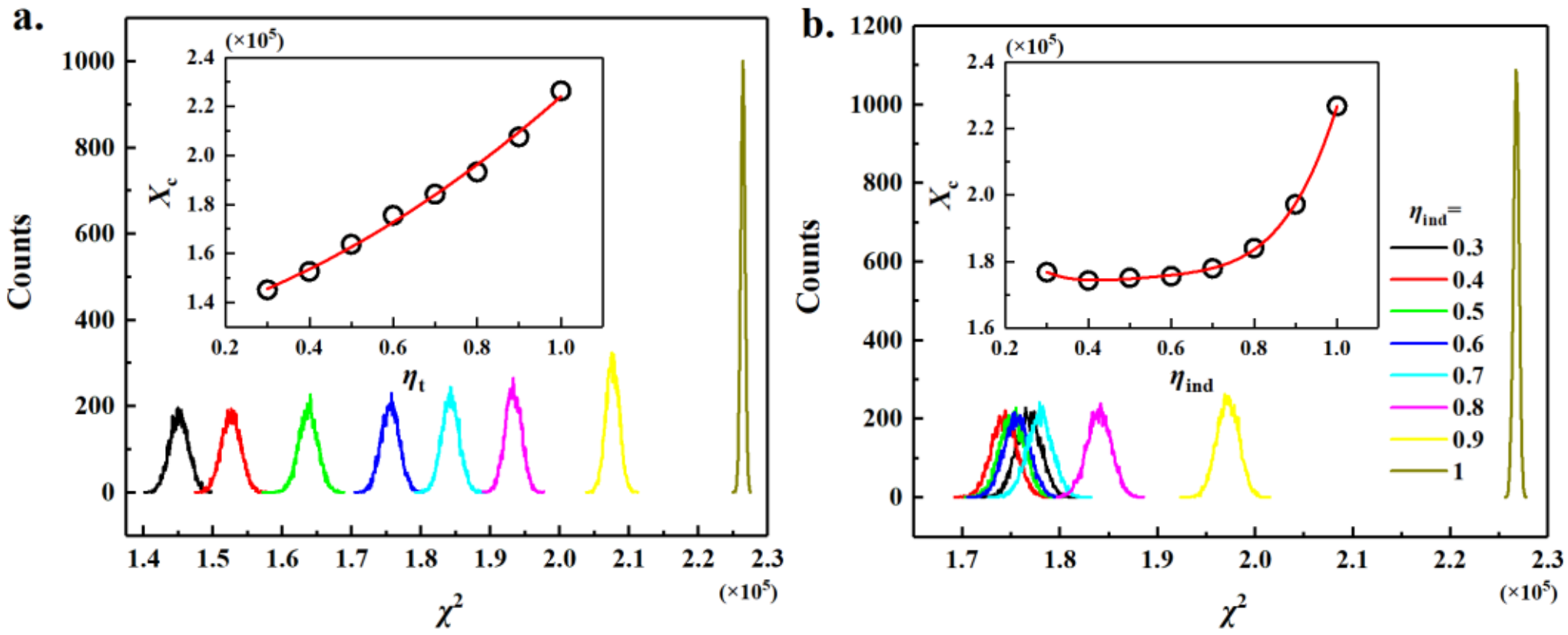

Fig. S4 Extended pattern recognition validations on noisy GBS including those in high-noise regimes. (a) Photon loss cases. (b) Photon distinguishability cases, with labels to distinguish crowed peaks. The basic sample and validation parameters are kept the same as those in small-scale cases in the manuscript.

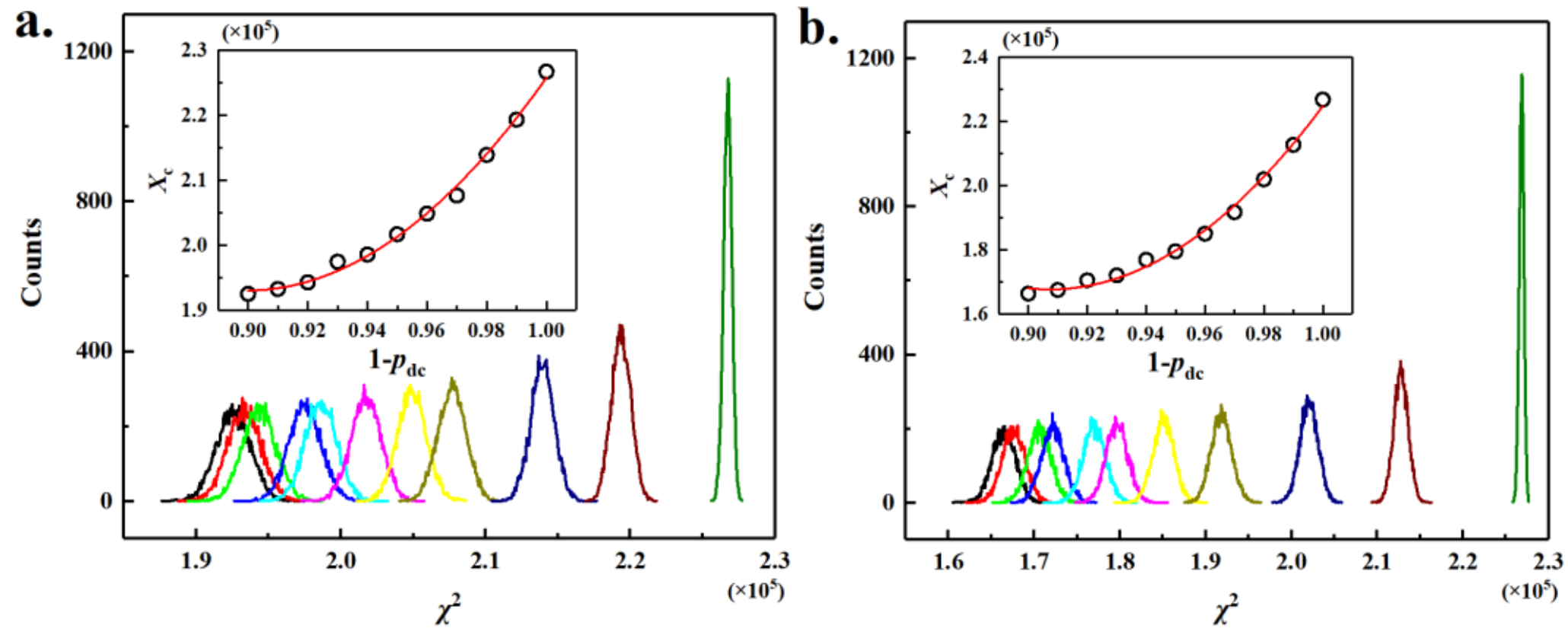

Fig. S5 Extended pattern recognition validations on GBS considering dark counts. (a) Cases with only the noise of dark counts. (b) Cases with noises of both photon loss and dark counts, where for each sample set, the transmission rate and the dark count probability are correlated by $\eta_{\mathrm{t}} = 1 - p_{\mathrm{dc}}$. The basic sample and validation parameters are kept the same as those in small-scale cases in the manuscript.

# 6 Validations on large-scale GBS

The lack of ideal samples restricts the validation approach to be well extended to evaluate noise levels, when the scale is large enough. On this condition, it may be up to those efficient classical approximation algorithms to generate bona fide samples. In order to make sure that the extended validation approach can still work, these approximate samples must be close enough to the ideal ones but different from those physically noisy samples.

Employing approximate samples as the bona fide ones is useful in a degree, which is partly supported by the results of our previous work focusing on extending pattern recognition validations on noisy boson sampling [10]. In our previous work, an efficient classical algorithm based on cutoff interference photon numbers is used to generate approximate samples. The cluster structures formed by training these approximate samples as the bona fide ones are successful to achieve the goal of evaluating noise levels of boson sampling. The cost is the suppression of the Gaussian peak center monotonicity with noises.

Here, we present a large-scale example of the extended validation on lossy GBS, running a classical algorithm proposed by C. Oh et al. [13] to generate approximate samples as the bona fide ones, with the assistance of the matrix product state (MPS) method. Please note that those lossy samples with high noises to be evaluated can also be exactly generated with a similar scheme. The only difference is that compared to that of the approximate samples, generating those exactly simulated lossy samples actually does not need the MPS method. This is classically feasible thanks to smaller expected total output photon numbers correlated to quantum processes and resulted from a smaller equivalent squeezing parameter.

For simplification, we do not introduce the approximate method, that is, MPS, in detail. However, we think it may be helpful for understanding, by presenting the efficient exact simulation method following the scheme proposed by C. Oh et al. [13] but replacing the MPS process with the exact simulation process mentioned in the manuscript, where the equivalent squeezing parameter we need to consider is obviously smaller than the original one. Please note that a smaller squeezing parameter means a smaller expected total output photon number. As a result, the computations can be simplified. Generally, the covariance matrix of the output state in lossy GBS is [13]

$$\boldsymbol{V} = \eta_{\mathrm{t}}\boldsymbol{V}_0 + (1-\eta_{\mathrm{t}})\boldsymbol{I}, \tag{S9}$$

where $\boldsymbol{V}_0$ corresponds to the output state without loss and $\eta_{\mathrm{t}}$ is the transmission rate. By reconstructing the right side of the equal sign, $\boldsymbol{V}$ can be rewritten as [13]

$$\boldsymbol{V} = \boldsymbol{V}_{\mathrm{p}} + \boldsymbol{W}, \tag{S10}$$

where $\boldsymbol{V}_{\mathrm{p}}$ is a covariance matrix of a pure state with a smaller equivalent squeezing parameter and $\boldsymbol{W}$ is a matrix corresponding to classical displacement parts. The equivalent squeezing parameter should be small enough to greatly simplify the MPS process or the exact simulation process. The benchmark is that the trace of $\boldsymbol{V}_{\mathrm{p}}$ should be small enough while the matrices $\boldsymbol{W}$ and $\boldsymbol{V}_{\mathrm{p}} - i\boldsymbol{\Omega}$ should be semi-positive definite to meet their physical requirements, where $\boldsymbol{\Omega}$ is a symplectic form. With the equivalent squeezing parameter obtained by considering the benchmark, the simulation method mentioned in the manuscript can be employed to generate temporary samples owing to $\boldsymbol{V}_{\mathrm{p}}$.

Once a temporary sample is obtained, a displacement operator is performed on the state locally, to get a final sample. The displacement operations are introduced relatively briefly in the reference [13]. Here, we provide a mathematical procedure in detail， partly following the route proposed in the reference [14]. The displacement is generated from a multivariate Gaussian distribution according to $\boldsymbol{W}$ [5], that is,

$$\Pr(\boldsymbol{\beta}) = \frac{\exp\left(-\frac{1}{2}\boldsymbol{\beta}^{\mathrm{t}}\boldsymbol{W}^{-1}\boldsymbol{\beta}\right)}{\sqrt{\det(2\pi\boldsymbol{W})}}. \tag{S11}$$

According to the decomposition theorem, the displacement operator on the Fock state $|n\rangle$ in the $i$th mode ($i = 1, \dots, m$, where $m$ is the total mode number) is [14]

$$\widehat{D}(\beta_i) = e^{-\frac{|\beta_i|^2}{2}} e^{\beta_i \hat{a}^\dagger} e^{-\beta_i^* \hat{a}}. \tag{S12}$$

Hence, we have

$$\widehat{D}(\beta_i)|n\rangle = e^{-\frac{|\beta_i|^2}{2}} e^{\beta_i \hat{a}^\dagger} e^{-\beta_i^* \hat{a}} |n\rangle \tag{S13}$$

$$= e^{-\frac{|\beta_i|^2}{2}} \sum_{k=0}^{\infty} \frac{\beta_i^{\,k}}{k!} \hat{a}^{\dagger k} \sum_{l=0}^{\infty} \frac{(-\beta_i^*)^l}{l!} \hat{a}^l \,|n\rangle \tag{S14}$$

$$= e^{-\frac{|\beta_i|^2}{2}} \sum_{k=0}^{\infty} \frac{\beta_i^{\,k}}{k!} \hat{a}^{\dagger k} \sum_{l=0}^{n} \frac{(-\beta_i^*)^l}{l!} \hat{a}^l \,|n\rangle. \tag{S15}$$

Considering the properties of creation and annihilation operators, to make sure that $\langle n'|\widehat{D}(\beta_i)|n\rangle$ is not zero, the resultant photon number $n'$ should meet the condition

$$n' = n + k - l. \tag{S16}$$

Hence, we have

$$\langle n'|\widehat{D}(\beta_i)|n\rangle = \langle n'|e^{-\frac{|\beta_i|^2}{2}} \sum_{\substack{l \geq n-n' \\ 0 \leq l \leq n}} \frac{\beta_i^{\,n'-n+l}(-\beta_i^*)^l}{(n'-n+l)!\,l!} \hat{a}^{\dagger\, n'-n+l} \hat{a}^l |n\rangle \tag{S17}$$

$$= \langle n'|e^{-\frac{|\beta_i|^2}{2}} \sum_{\substack{l \geq n-n' \\ 0 \leq l \leq n}} \frac{\beta_i^{\,n'-n+l}(-\beta_i^*)^l}{(n'-n+l)!\,l!} \left(\prod_{j=0}^{l-1}\sqrt{n-j}\right) \hat{a}^{\dagger\, n'-n+l} |n-l\rangle \tag{S18}$$

$$= e^{-\frac{|\beta_i|^2}{2}} \sum_{\substack{l \geq n-n' \\ 0 \leq l \leq n}} \frac{\beta_i^{\,n'-n+l}(-\beta_i^*)^l}{(n'-n+l)!\,l!} \left(\prod_{j=0}^{l-1}\sqrt{n-j}\right)\left(\prod_{j'=0}^{n'-n+l-1}\sqrt{n'-j'}\right). \tag{S19}$$

Based on this, we can get the final output number $n'$ on the mode according to the probabilities written as $\left|\langle n'|\widehat{D}(\beta_i)|n\rangle\right|^2$. These operations are local, which means that the computations are fast.

In Fig. S6, we present two small-scale lossy examples which support the idea that our simulations are reliable. To compare with those samples exactly simulated with the method mentioned above (Sample Set No. 2) or approximately simulated with the MPS method (Sample Set No. 1), a series of exactly simulated samples with the method mentioned in the manuscript (Sample Set No. 0) are also generated. Compared to Sample Set No. 0, Sample Set No. 1 has a total variation distance (TVD) of 0.1025 and a similarity of 0.9918 while Sample Set No. 2 has a TVD of 0.1021 and a similarity of 0.9924. These results indicate that the exactly simulated samples here are closer to the theoretic lossy ones.

In Fig. S7, the total output photon numbers of large-scale approximate samples to be used as the bona fide ones are counted, which can reach up to several dozen and thus are comparable to those of the early-reported experiments [13].

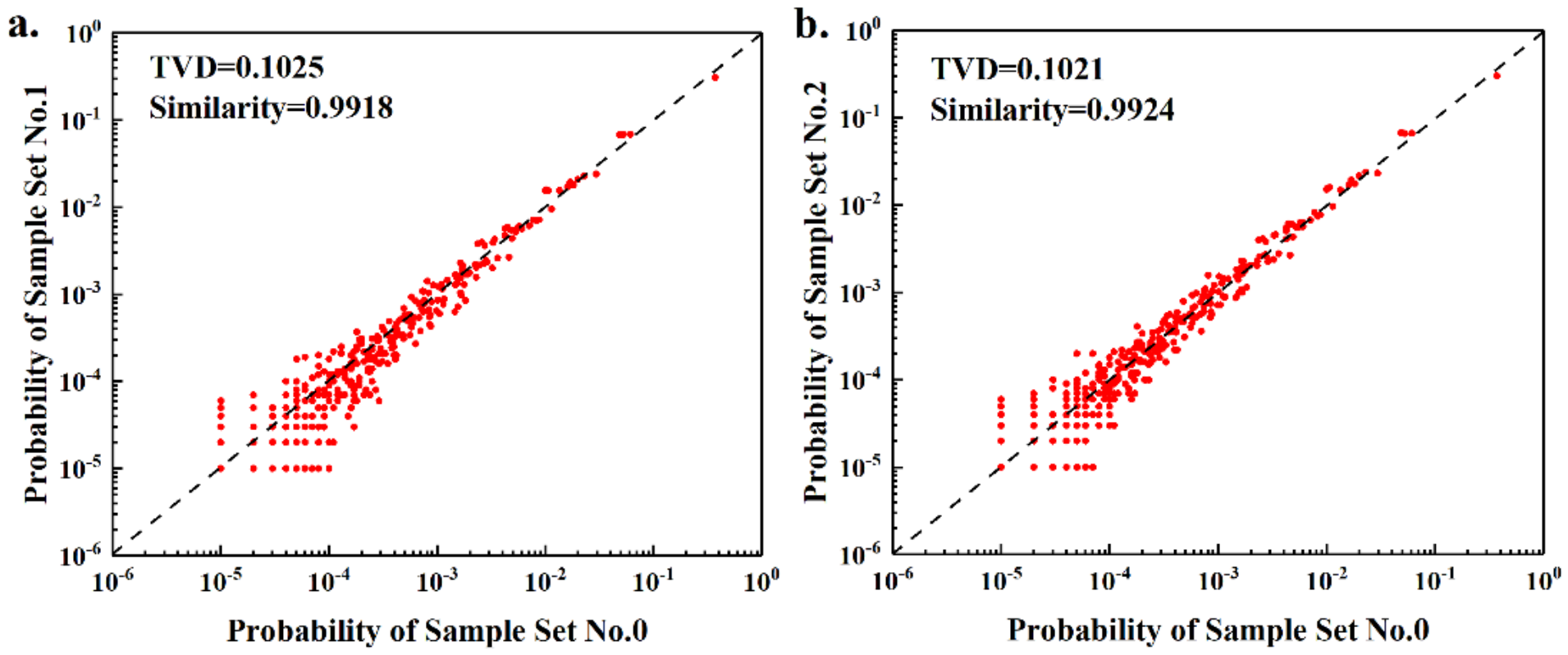

Fig. S6 Probabilities of different small-scale sample sets. (a) Comparisons between samples exactly simulated with the method mentioned in the manuscript and samples simulated by MPS, with $m = K = 4$, $r = 1$, $\eta_t = 0.3$, $n_{\text{cutoff}} = 4$. (b) Comparisons between two kinds of exactly simulated samples. The total sample number for each kind of sample sets is $10^5$. During the MPS simulation, the bond dimension reflecting the degree of entanglements is set as 10.

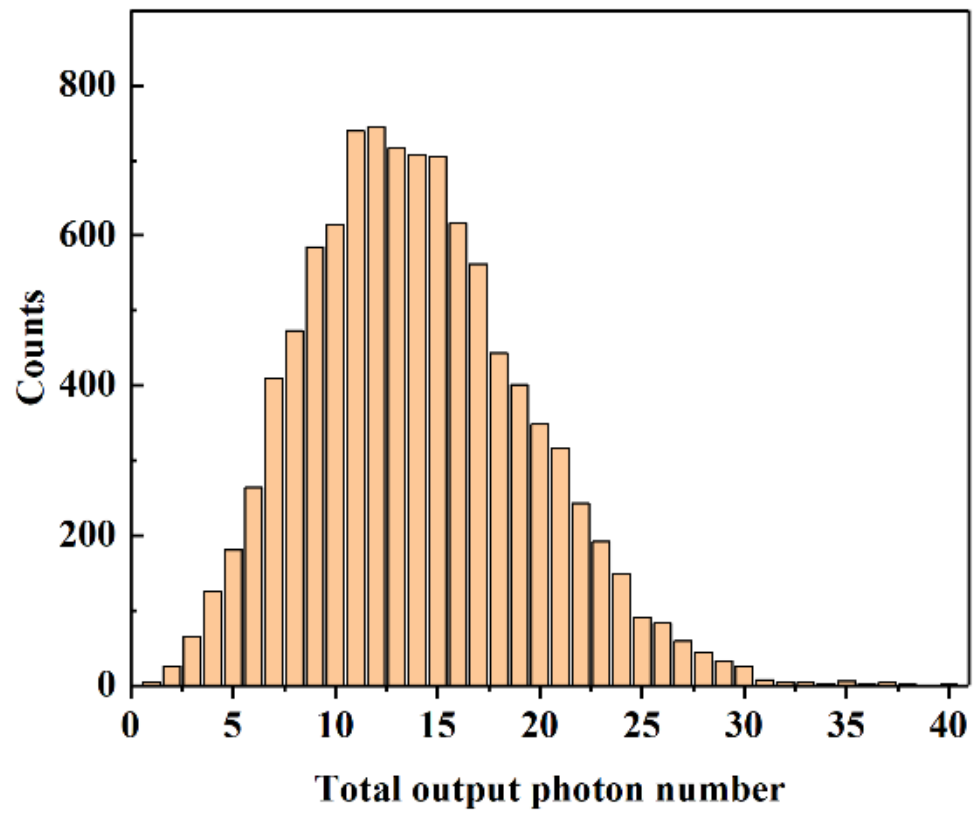

Fig. S7 Photon number statistics of approximate samples simulated by MPS, with $m = K = 20$, $r = 1$, $\eta_t = 0.7$, $n_{\text{cutoff}} = 5$. To simplify computations, the bond dimension is only set as 10, which means that the approximation is relatively rough. The total sample number is $10^4$.

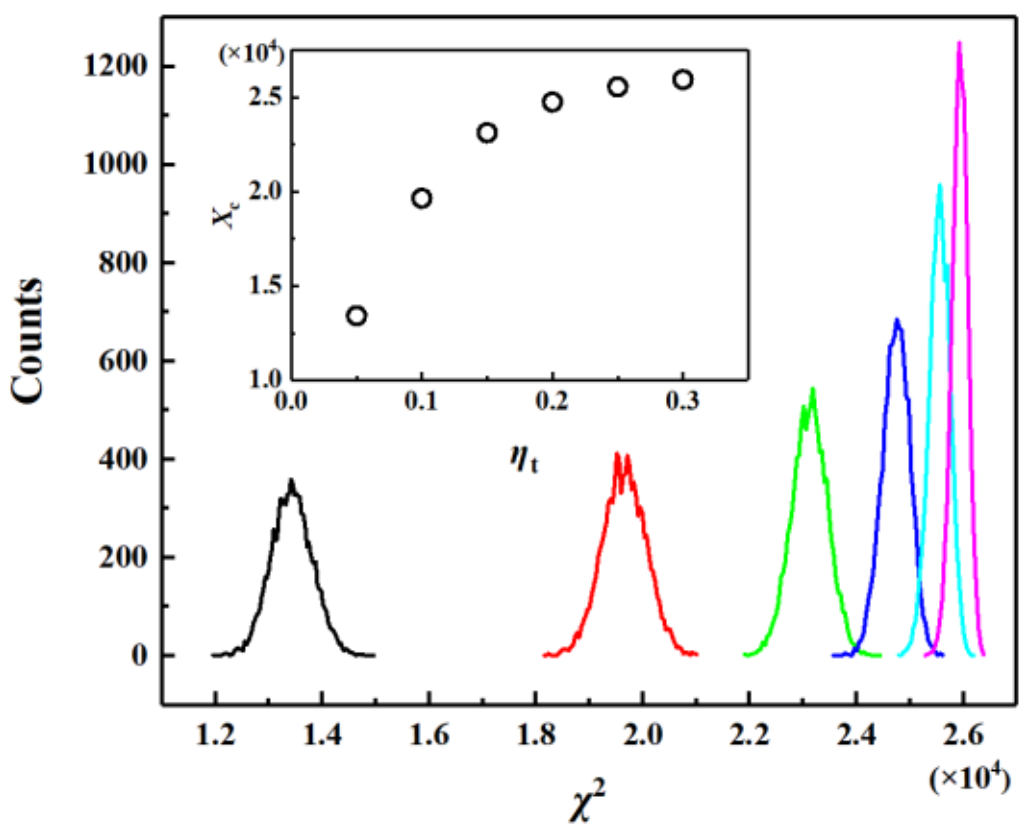

Fig. S8 Extended pattern recognition validations on exactly simulated large-scale lossy GBS samples, where approximate samples generated by MPS are employed as the bona fide ones. For each kind of samples, the total number is $10^4$, with $m = K = 20$, $r = 1$, $n_{\text{cutoff}} = 5$. The cluster number is 50 and the training sample number is 1000, where every 4 adjacent modes are bound together as a subset to employ the output-binning strategy. The inset shows the relationship of the Gaussian peak center $X_{\text{c}}$ and the transmission rate $\eta_{\text{t}}$.

With clusters constructed by training the approximate samples, the noise levels of those exactly simulated samples with high loss can be evaluated based on the extended pattern recognition validation approach combined with the output-binning strategy mentioned in the manuscript. As shown in Fig. S8, the Gaussian peak center location is still monotonic with loss while the tendency is not in accordance with those in the manuscript. This is perhaps because the clusters are coarse considering the finite samples. Moreover, even if the clusters are robust and characteristic, which well reflecting the data structures of the bona fide samples, Gaussian peaks may still keep close whenever similarities between those noisy samples and the bona fide samples which are generated approximately are changeless. Therefore, in order to get a good extended validation performance, the approximate simulations should be reliable enough to make sure that the approximate samples are different from those noisy ones. In MPS, for example, an effective route may be increasing the bond dimension, which, however, means computational cost increasements.

On the other hand, if the total number of modes increases while the expected output photon number is fixed, the validation approach can perform efficiently more or less, owing to the output-binning strategy. Here, we provide an example to support this idea, where the mode number $K$ with single-mode squeezed states as the input and the squeezing parameter $r$ are fixed, which means a fixed expected output photon number, while the total mode number $m$ is changed. The total sample number is also fixed. As shown in Fig. S9, when the mode number is increased, changes of Gaussian peak center locations are unobvious, where the validation parameters are kept the same, except that the output-binning strategy is employed in the case with a larger mode number. This example indicates that with increasing the mode number while keeping the photon number fixed, the extended validation approach may still work efficiently owing to the output-binning strategy.

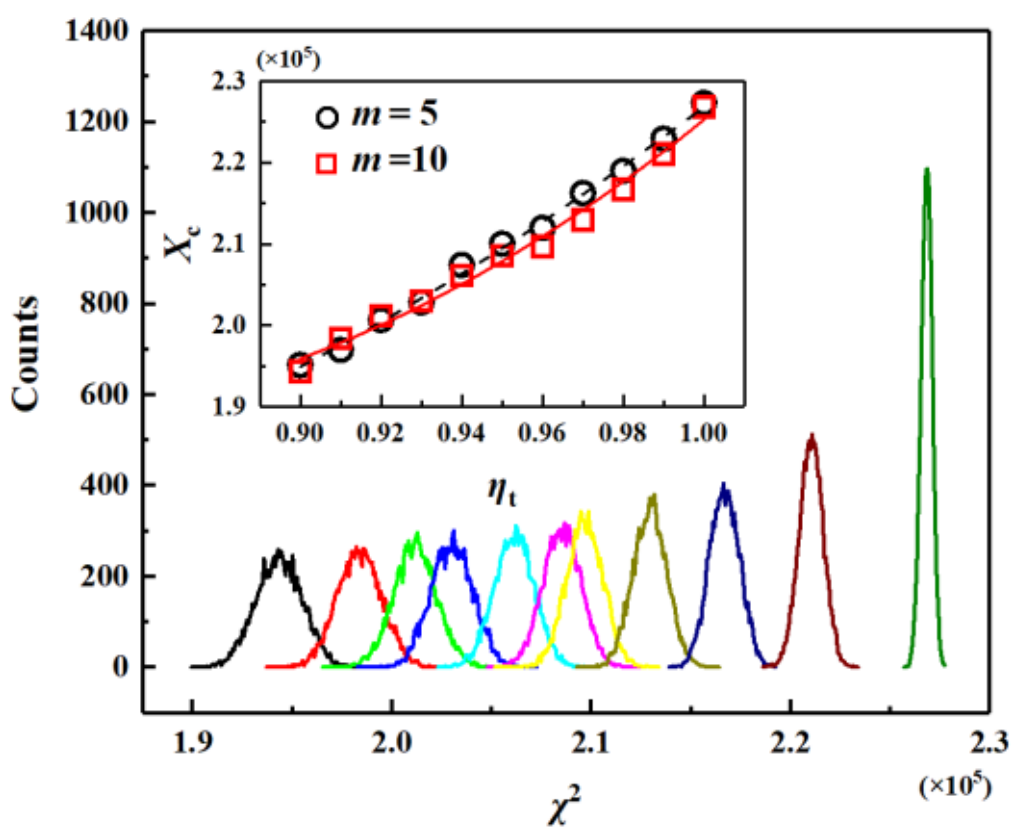

Fig. S9 Extended pattern recognition validations on lossy GBS with a larger mode number $m$, which is set as 10. The other basic sample parameters are the same as those of the small-scale lossy samples in the manuscript. The validation parameters are also kept the same. The output-binning strategy is used where the adjacent modes are bound together as a subset. In the inset, the Gaussian peak center locations corresponding to the small-scale lossy samples in the manuscript with $m = 5$ are also presented for comparisons. The black dashed and red solid lines are fitted curves corresponding to $m = 5$ and $m = 10$, respectively.

# 7 Validation approach comparisons with finite samples

We make comparisons between pattern recognition approaches employing the output-binning strategy, and correlation approaches, on GBS samples simultaneously influenced by noises of both photon loss and photon distinguishability. The sample numbers are restricted. With respect to the 1st- to 4th-order correlators, the characteristic value $\Gamma$ shown in Fig. S10 is in an increasing tendency when noises become lower. However, there are some points going against the total tendency due to sample finiteness. As a contrast, the pattern recognition approach performs better, as shown in Fig. S11.

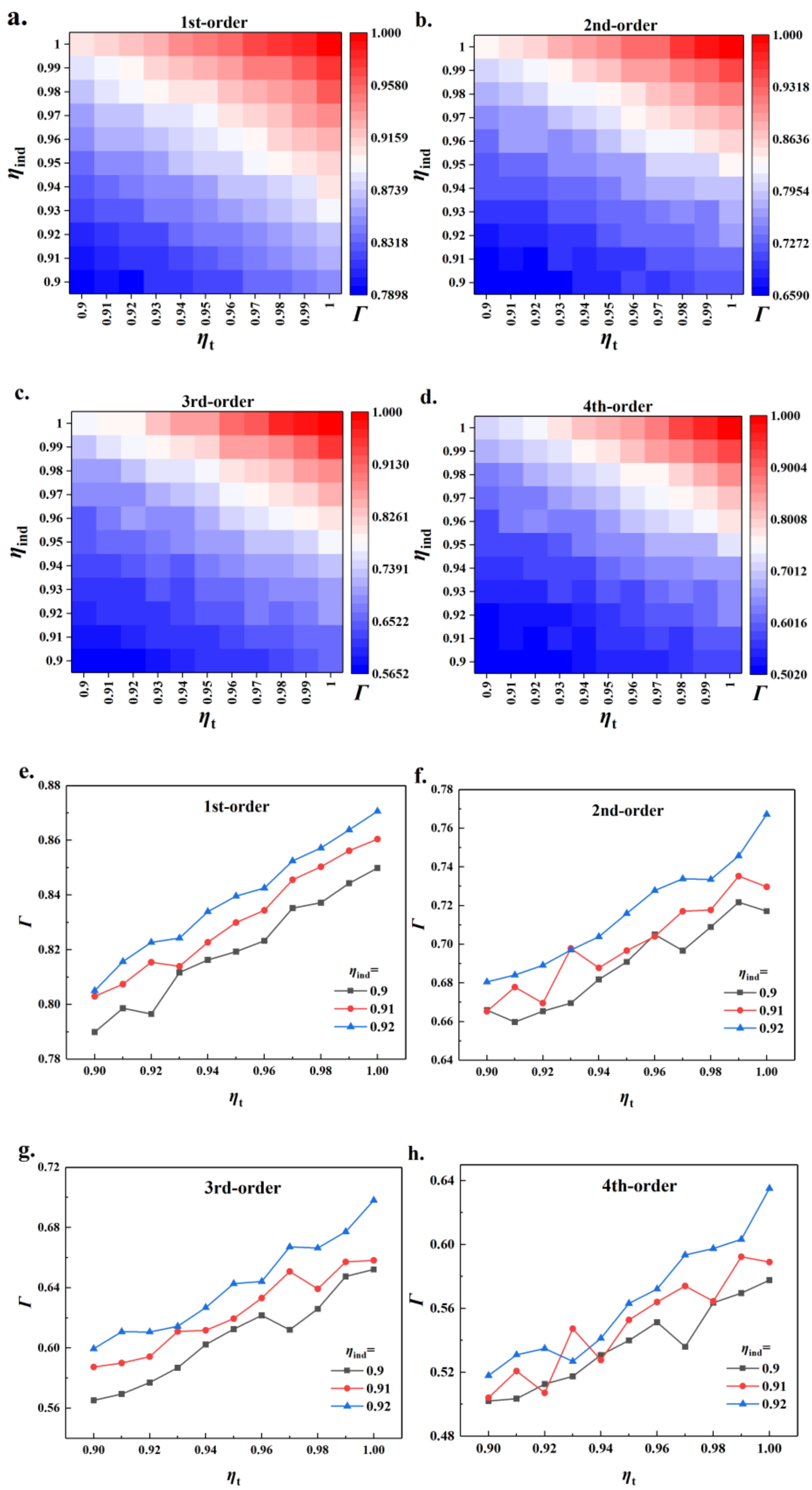

Fig. S10 Correlation validation performances on GBS samples influenced by the two main kind noises. (a)-(d) $\Gamma$ maps of 1st- to 4th-order correlators, respectively. (e)-(h) Corresponding plots of $\Gamma$ and $\eta_{\rm t}$ with $\eta_{\rm ind}$ kept as 0.9, 0.91 and 0.92. The basic parameters of GBS samples are the same as those in larger-scale cases in the manuscript.

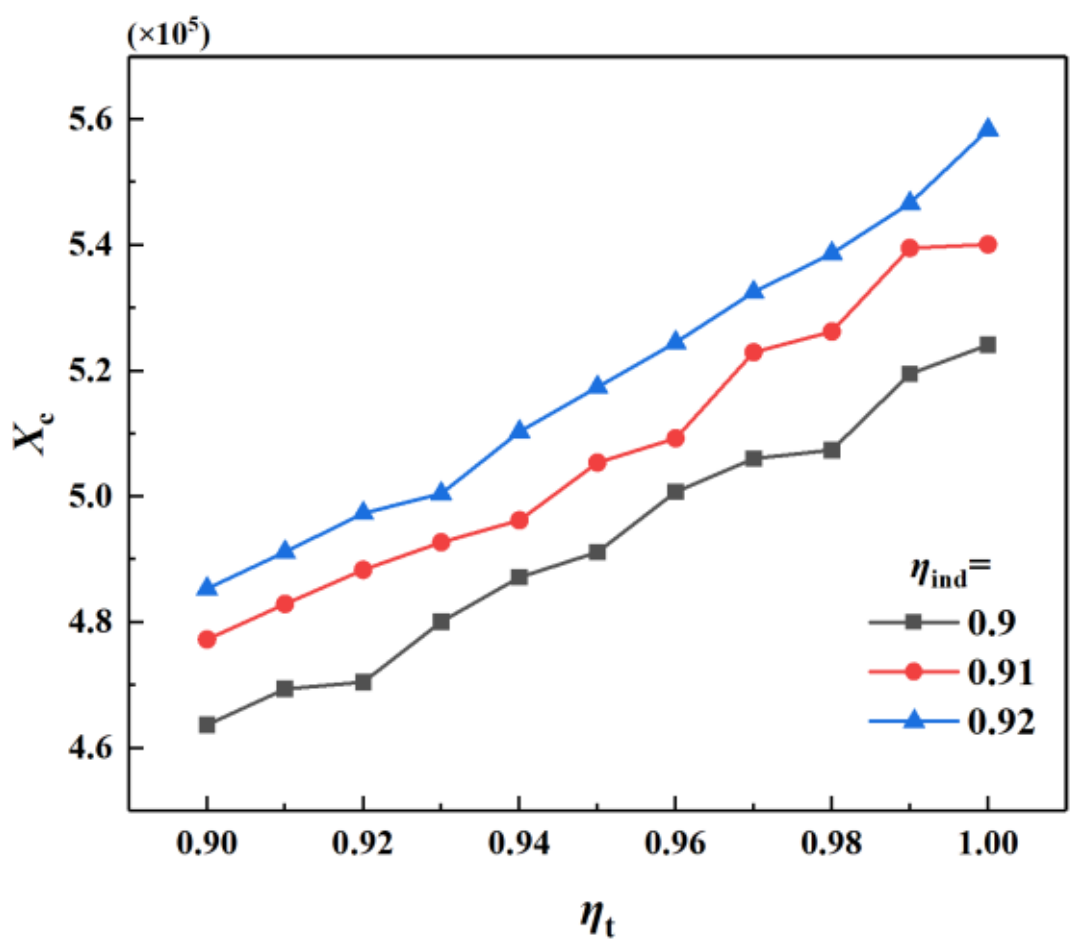

Fig. S11 Plots of $X_c$ and $\eta_t$ with $\eta_{ind}$ kept as 0.9, 0.91 and 0.92. The data are extracted from Fig. 4 in the manuscript.